\documentclass[pdflatex,sn-aps,a4paper]{sn-jnl}
\usepackage{amsmath,amssymb,graphicx,color}
\usepackage{bm}
\usepackage{bbm}
\usepackage[usenames,dvipsnames]{xcolor}
\usepackage[normalem]{ulem}
\usepackage{soul,xcolor}
\usepackage{subfigure}

\newcommand{\avg}[1]{\ensuremath{\langle #1 \rangle}}

\usepackage{fixltx2e,amsmath}
\MakeRobust{\eqref}

\newcommand{\ve}[1]{\bm{#1}}
\usepackage{mathptmx} 

\begin{document}
\setstcolor{blue}
\title{Universal relations and bounds for fluctuations in quasistatic small heat engines}
\author[1,2,3,10]{Kosuke Ito}
\author[1,10]{Guo-Hua Xu}
\author[1,4]{Chao Jiang}
\author[5]{\'Edgar Rold\'an}
\author[6,7]{Ra\'ul~A. Rica-Alarc\'on}
\author[8]{Ignacio~A. Mart\'inez}
\author*[1,9]{Gentaro Watanabe}\email{gentaro@zju.edu.cn}
\affil[1]{School of Physics and Zhejiang Institute of Modern Physics, Zhejiang University, Hangzhou, Zhejiang 310027, China}
\affil[2]{Graduate School of Engineering Science, Osaka University, 1-3 Machikaneyama, Toyonaka, Osaka 560-8531, Japan}
\affil[3]{Center for Quantum Information and Quantum Biology, Osaka University, 1-2 Machikaneyama, Toyonaka, Osaka 560-8531, Japan}
\affil[4]{Graduate School of China Academy of Engineering Physics, Beijing 100193, China}
\affil[5]{ICTP -- The Abdus Salam International Centre for Theoretical Physics, Strada Costiera 11, 34151 Trieste, Italy}
\affil[6]{Universidad de Granada, Department of Applied Physics, 18071 Granada, Spain}
\affil[7]{Universidad de Granada, Nanoparticles Trapping Laboratory and Research Unit ``Modeling Nature'' (MNat), 18071 Granada, Spain}
\affil[8]{Department of Electronics and Information Systems, Ghent University, Technologiepark-Zwijnaarde 126, 9052 Ghent, Belgium}
\affil[9]{Zhejiang Province Key Laboratory of Quantum Technology and Device, Zhejiang University, Hangzhou, Zhejiang 310027, China}
\affil[10]{These authors contributed equally: Kosuke Ito, Guo-Hua Xu}

\date{\today}

\abstract{The efficiency of any heat engine, defined as the ratio of average work output to heat input, is bounded by Carnot's celebrated result. However, this measure is insufficient to characterize the properties of miniaturized heat engines carrying non-negligible fluctuations, and a study of higher-order statistics of their energy exchanges is required. Here, we generalize Carnot's result for reversible cycles to arbitrary order moment of the work and heat fluctuations. Our results show that, in the quasistatic limit, higher-order statistics of a small engine's energetics depend solely on the ratio between the temperatures of the thermal baths. We further prove that our result for the second moment gives universal bounds for the ratio between the variances of work and heat for quasistatic cycles. We test this theory with our previous experimental results of a Brownian Carnot engine and observe the consistency between them, even beyond the quasistatic regime. Our results can be exploited in the design of thermal nanomachines to reduce their fluctuations of work output without marginalizing its average value and efficiency.}


\maketitle

\bigskip
\noindent \textbf{\textsf{\large Introduction}}

Formidable advances in physics and nanotechnology have led to the construction of miniaturized heat engines of the size of a colloidal particle~\cite{Hugel02,Blickle12,Quinto-Su14,Martinez16,Krishnamurthy16,Martinez17,Argun17} or even smaller~\cite{Steeneken11,Koski14,Rossnagel16,Maslennikov19,Klatzow19,Lindenfels19,Peterson18}. Such microscopic heat engines constitute a perfect target study for understanding the double-sided challenges of miniaturization. On the one side, there is the aim at understanding of the fundamental thermodynamic principles of small machines~\cite{Sekimoto98,Sekimotobook10,Jarzynski11,Seifert12,VandenBroeck14,Ciliberto17,Nicolis17}; on the other side, there is the pursuit for the optimal design of artificial devices~\cite{Bustamante05,Seifert12,Toyabe15,Martinez17,Fodor21}. The relevance of fluctuations in thermodynamic quantities of small machines such as work, heat or entropy has been predicted by the theoretical framework of stochastic thermodynamics \cite{Sekimoto00,Jop08,Hoppenau13,Holubec14,Rana14,Cerino15,Holubec17,Miller20,Watanabe21,Xu21}, and reported in numerous experimental realizations of mesoscopic heat engines with tremendous accuracy~\cite{Blickle12,Martinez15,Martinez16,Krishnamurthy16}. As demonstrated by the experiment of Ref.~\cite{Martinez16}, for instance, fluctuations of thermodynamic quantities can be dominant over their average values. Such theoretical and experimental progress highlighted the need to characterize small machines' fluctuations through higher moments of their thermodynamic quantities beyond their mean values~\cite{Verley14a,Verley14b,Polettini15,Proesmans15a,Jiang15,Proesmans15b,Vroylandt16,Park16,Basu17,Gupta17,Saha18,Sune19,Manikandan19}. Although Carnot's celebrated result regarding the ratio between average work and heat of cycles has been known for a long time, there is still no universal bound on the fluctuations of work and heat, even for quasistatic cycles.

In this article, we develop a theory that describes the $(n\geq 2)$-th order moments of the work and heat of small heat engines that are subject to thermal fluctuations and test our predictions with experimental data from the Brownian Carnot engine~\cite{Martinez15}, a Carnot engine with a Brownian particle as its working substance. In particular, we analyze the statistical properties of reversible Carnot cycles through moments of arbitrary order, finding that they show universal features in the line of Carnot's classical result: with exclusive dependence on the temperature ratio $T_{\rm c}/T_{\rm h}$ between the hot $(T_{\rm h})$ and cold $(T_{\rm c})$ heat baths. Furthermore, we show that the ratio between the fluctuations of work and heat for the Carnot cycle provides a universal upper bound for this quantity among cycles operating between the temperatures $T_{\rm h}$ and $T_{\rm c}$ consisting of quasistatic strokes. Our analytical results are tested using previous experimental data of the Brownian Carnot engine~\cite{Martinez15}, exploring their validity from the quasistatic limit to the non-equilibrium performance in the non-quasistatic regime. Our results pave the way to designing reliable microscopic heat engines by reducing the fluctuations of the work output without lowering the average work and efficiency.

\bigskip
\noindent \textbf{\textsf{\large Results}}

\noindent \textbf{Working substance}

Systems with a small number of degrees of freedom do not necessarily relax into an equilibrium state during adiabatic processes even if they are quasistatic (Throughout the present paper, the word ``adiabatic'' means that there is no energy exchange between the system and environment in the form of heat.). For example, systems with more than two discrete energy levels initially in equilibrium at some temperature is no longer equilibrium for any temperature after the energy of one of the levels is changed quasistatically. Therefore, in establishing thermal contact with a heat bath after an adiabatic process, an irreversible change in the energy distribution occurs if the final state of the adiabatic process does not follow the canonical distribution at the temperature of the bath \cite{Sekimotobook10}. To realize the reversible Carnot cycle with a small system, an extra condition is required to avoid such irreversibility, which does not exist for macroscopic systems. In the present work, we consider a class of working substances that allows us to implement the desired cycle free from this irreversibility for arbitrary values of $T_{\rm h}$ and $T_{\rm c}$. For the experimental platform to realize our setup, systems with high controllability and high (and controllable) isolation from the environment such as trapped ions~\cite{Leibfried03} and levitated nanoparticles~\cite{Gonzalez-Ballestero21,Winstone23} are promising.

Suppose a system is in thermal equilibrium at temperature $T_1$ with an external control parameter $\lambda$ at $\lambda_1$. Then, a quasistatic adiabatic process is performed by slowly changing $\lambda$ from $\lambda_1$ to $\lambda_2$. After the quasistatic adiabatic process, the system is put in contact with a thermal bath at temperature $T_2$. The internal energy of the initial and final state of the adiabatic process is denoted by $E_1$ and $E_2$, respectively. At the beginning of the quasistatic adiabatic process, the system is in a microstate $\Gamma_1$ in the phase space with energy $E_1 = H_{\lambda_1}(\Gamma_1)$ whose distribution is given by the canonical ensemble for $H_{\lambda_1}$ at $T_1$. During the quasistatic adiabatic process, the system evolves deterministically by Hamilton's equations of motion and the internal energy of the system changes following the adiabatic theorem such that the phase space volume $I_\lambda(E)$ enclosed by the iso-energy surface at $E$ (the so-called number of states) is invariant under the slow change of $\lambda$. Here, $I_\lambda(E) \equiv \int \theta\left(E - H_{\lambda}(\Gamma)\right)\, d\Gamma$ with $\theta$ being the Heaviside step function. To ensure reversibility, the energy distribution before and after the thermal contact should be the same, so 
\begin{align}
  \frac{E_1}{T_1} = \frac{E_2}{T_2}\nonumber
\end{align}
up to a constant for any realization [adiabatic-reversibility (AR) condition] \cite{Sato02,Sekimotobook10}. In order to guarantee that we can find an appropriate final parameter $\lambda_2$ that fulfills the AR condition, the initial and final energies of the quasistatic adiabatic process need to be linked by $E_1 = \phi(\lambda_1, \lambda_2)\, E_2$ with a function $\phi(\lambda_1,\lambda_2)$ that solely depends on the initial and final parameter values $\lambda_1$ and $\lambda_2$~\cite{Sato02}.

To implement the reversible Carnot cycle for any choice of the baths' temperatures and the parameter $\lambda$ at the starting (or ending) point of the adiabatic strokes, the AR condition has to be satisfied for arbitrary values of $T_1$, $T_2$, and $\lambda_1$ (or $\lambda_2$).
Hence, the working substance should satisfy the above relation, $E_1 = \phi(\lambda_1, \lambda_2)\, E_2$, with $\phi(\lambda_1,\lambda_2)$ such that for every fixed $\lambda_1=\lambda$, the function $g_{\lambda}(\lambda_2)\equiv \phi(\lambda,\lambda_2)$ of $\lambda_2$ can take any value of $T_1/T_2$. It is noted that the typical working substance for microscopic heat engines falls in this class of working substance. For example, a single particle trapped in a harmonic oscillator potential with controllable spring constant, which is commonly employed in most of the experiments so far~\cite{Blickle12,Martinez15,Martinez16,Krishnamurthy16,Argun17}, and a single particle trapped in a box potential with controllable box width are the cases since their number of states is in the form of $I_{\lambda}(E) = f(\lambda)\,E^{\alpha}$ with a positive unbounded function~$f$. This holds, e.g., for a scale-invariant potential $V_\lambda(x)$ satisfying $V_{\lambda}(ax) = |a|^b\, V_{\lambda}(x)$ with positive real constant $b>0$, where $a$ is the scaling factor and $x$ represents the position of the particle.

\begin{figure}[t!]
\centering
\includegraphics[width=1 \columnwidth]{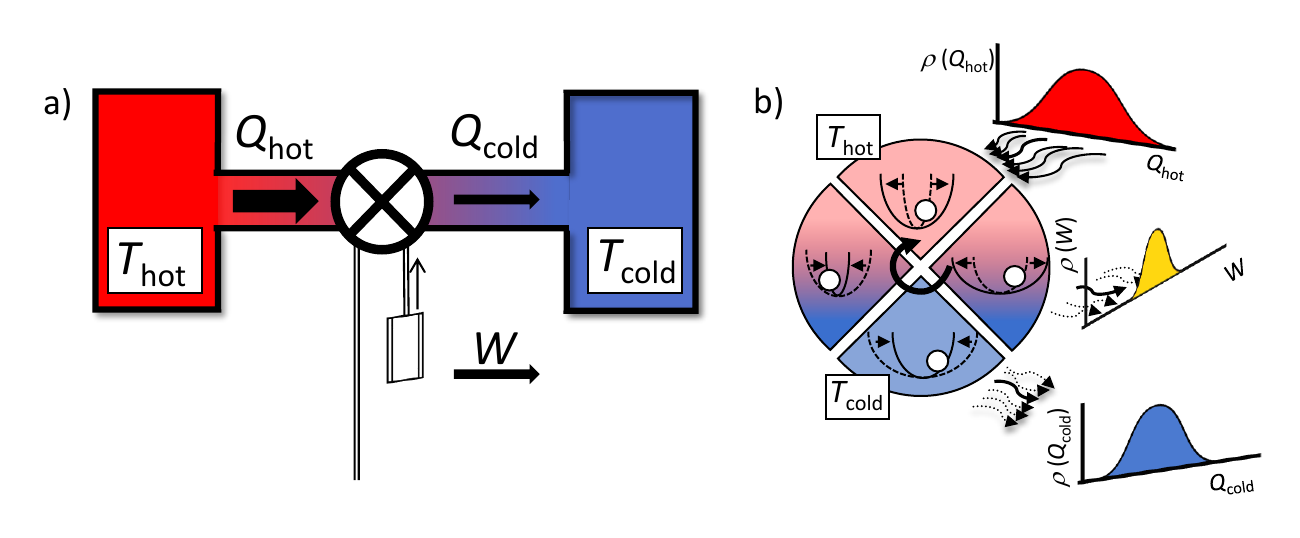}
\caption{\textbf{Schematics of a Brownian Carnot engine.} (a) Carnot's theorem bounds the maximum work that can be extracted from two thermal baths based on the ratio between the hot and cold temperatures. (b) When the heat engine is miniaturized, see standard protocol depicted between the hot bath (up, red) and the cold one (down, blue) with the four processes sketched, thermal fluctuations become significant, making all the energy fluxes stochastic, described by their probability density functions. Our results show that the various moments of the energetics also scale with the temperature ratio.}
\label{fig:carnot}
\end{figure}

\bigskip
\noindent \textbf{Universal relations between work and heat fluctuations of the Carnot cycle}

All relevant quantities of a microscopic heat engine are stochastic by nature, and therefore they fluctuate over different realizations of a given protocol. When averaged over many realizations of a given protocol, the mean values of work and heat converge, and their stochastic nature is revealed by their probability distributions [Fig.~\ref{fig:carnot}(b)].

In many stochastic systems, mean values are not good quantities to characterize their behavior, but the moments of their distributions need to be accounted for. It is indeed the case for the experiments of the Brownian heat engines, where the fluctuations of thermodynamic quantities are comparable to their mean values \cite{Blickle12,Martinez16}. The $n$-th order central moment $\avg{(\Delta X)^n}$ of a random variable $X$ is defined as $\avg{(\Delta X)^n} \equiv \avg{(X - \avg{X})^n}$, where $\avg{X}$ and $\Delta X \equiv X - \avg{X}$ are the statistical average and the fluctuation of ``$X$'', respectively. As we detail in the subsection of ``Derivation of Eqs.~(\ref{eq:etan}), (\ref{eq:nthcentralrel}), and (\ref{eq:xi2})'' in the ``Methods'' section, one can derive a neat relation for the higher-order statistics of the work output $W$ and the heat input $Q_{\rm h}$ for the reversible Carnot cycle:
\begin{align}
  \eta^{(n)} \equiv \frac{\avg{(\Delta W)^n}}{\avg{(\Delta Q_{\rm h})^n}} = \left( 1 - \frac{T_{\rm c}}{T_{\rm h}} \right)^n \equiv \eta_{\rm C}^n\,,\label{eq:etan}
\end{align}
which holds for any integer $n \ge 2$. Here, $T_{\rm h}$ and $T_{\rm c}$ are the temperatures of the hot and cold heat baths, respectively, and $\eta_{\rm C} \equiv 1-(T_{\rm c}/T_{\rm h})$ is the Carnot efficiency. This formula relates any higher-order statistics of work output and heat input such as variances for $n=2$, skewness for $n=3$, kurtosis for $n=4$, and so forth. Equation (\ref{eq:etan}) is one of the main results of this work. It is noted that the ratio $\eta^{(n)}$ for the Carnot cycle given by Eq.~(\ref{eq:etan}) shows the universal form which depends only on the ratio $T_{\rm c}/T_{\rm h}$. As we will demonstrate later, our relation (\ref{eq:etan}) is more useful to control each higher-order statistics of $W$ compared to the stochastic efficiency $\hat{\eta}\equiv W/Q_{\rm h}$ since Eq.~(\ref{eq:etan}) contains only the central moments of $W$ and $Q_{\rm h}$ of the same order. The relation (\ref{eq:etan}) is obtained from the AR condition and from the fact that the fluctuations of work through quasistatic isothermal strokes are negligible: the fluctuations of work output $W_{\rm isoth}$ through the isothermal stroke vanish in the limit of long duration $\tau$ of the stroke as $\Delta W_{\rm isoth} = W_{\rm isoth} - \avg{W_{\rm isoth}} = O(\tau^{-1/2})$~\cite{Sekimotobook10,Holubec18} (see also the subsection of ``Fluctuations of work and heat in the quasistatic isothermal process'' in the ``Methods'' section for details).

The second relation that we have obtained (see the subsection of ``Derivation of Eqs.~(\ref{eq:etan}), (\ref{eq:nthcentralrel}), and (\ref{eq:xi2})'' in the ``Methods'' section for details of the derivation) connects the central moments of the heat exchanged with the heat baths and their temperatures:
\begin{align}
  \frac{\avg{(\Delta Q_{\rm h})^n}}{T_{\rm h}^n} = \frac{\avg{(\Delta Q_{\rm c})^n}}{T_{\rm c}^n}\,\label{eq:nthcentralrel}
\end{align}
for any integer $n \ge 2$. It is noted that this relation is analogous to the ``central relation of thermodynamics'' called by Feynman~\cite{Feynmanlect}, $\avg{Q_{\rm h}}/T_{\rm h} = \avg{Q_{\rm c}}/T_{\rm c}$, which holds between the two isothermal strokes in the reversible Carnot cycle. Since the temperatures $T_{\rm c}$ and $T_{\rm h}$ are arbitrary, these relations hold for two arbitrary quasistatic isotherms connected by two quasistatic adiabats. Finally, the variance of work and the sum $\Delta Q^{(2)}$ of the heat fluctuations over the separated sequences of heat-exchanging strokes, given by 
\begin{align}
    \Delta Q^{(2)} \equiv& \avg{\Delta Q_{\rm h}^2} + \avg{\Delta Q_{\rm c}^2}
  = \left[ 1 + (T_{\rm c}/T_{\rm h})^2 \right] \avg{\Delta Q_{\rm h}^2},\,\label{eq:deltaq2}
\end{align}
satisfy another universal relation depending solely on the ratio $T_{\rm c}/T_{\rm h}$:
\begin{align}
  \xi^{(2)} \equiv \frac{\avg{\Delta W^2}}{\Delta Q^{(2)}} = \frac{\left[1- \left(T_{\rm c}/T_{\rm h}\right)\right]^2}{1+ (T_{\rm c}/T_{\rm h})^2 }\,.\label{eq:xi2}
\end{align}

Here, we remark that the same relation as Eq.~(\ref{eq:etan}) could also be derived from the integral fluctuation theorem~\cite{Jarzynski00,Pal17}. As discussed in detail in the subsection of ``Integral fluctuation theorems and $\eta_{\rm C}^{(n)}$'' in the ``Methods'' section, we could obtain $W = [1- (T_{\rm c}/T_{\rm h})] Q_{\rm h}$ from the integral fluctuation theorem for a cyclic protocol starting from the equilibrium state at $T_{\rm c}$. Similarly, for a protocol starting from the equilibrium state at $T_{\rm h}$, we obtain $W = [(T_{\rm h}/T_{\rm c}) -1 ] Q_{\rm c}$. However, it is noted that the random variables in the above two equations obtained from the fluctuation theorems for different initial points should be regarded as different random variables (even for $W$ in these equations), which belong to different ensembles. Obviously, the fluctuation $\Delta E$ of the internal energy change through one cycle, $E = (Q_{\rm h} - Q_{\rm c}) - W$, depends on the temperature of the initial point, so that work and heat for different protocols with different initial points are different random variables. Therefore, it is not trivial to derive the relations given by Eqs.~(\ref{eq:nthcentralrel}) and (\ref{eq:xi2}) by combining the above two equations obtained for different initial points, and also not trivial to prove $W = [1- (T_{\rm c}/T_{\rm h})] Q_{\rm h}$ for an initial state other than the equilibrium state at $T_{\rm c}$ from the integral fluctuation theorems. While it is hard to discuss $W$, $Q_{\rm h}$, and $Q_{\rm c}$ for the same protocol in the derivation based on the integral fluctuation theorems, our bottom-up approach employed in the subsection of ``Derivation of Eqs.~(\ref{eq:etan}), (\ref{eq:nthcentralrel}), and (\ref{eq:xi2})'' in the ``Methods'' section can handle all the random variables of $W$, $Q_{\rm h}$, and $Q_{\rm c}$ for the same protocol at the same time, and allows us to deal with the effects of the end points more carefully, which are important in the discussion of the fluctuations for a single cycle. In addition, for our bottom-up approach, deriving the relations (\ref{eq:etan}), (\ref{eq:nthcentralrel}), and (\ref{eq:xi2}) for any other initial point is straightforward. Because of the above reasons, we took the bottom-up approach in our derivation provided in the subsection of ``Derivation of Eqs.~(\ref{eq:etan}), (\ref{eq:nthcentralrel}), and (\ref{eq:xi2})'' in the ``Methods'' section.

\begin{figure}[t]
\centering
\includegraphics[width=0.96 \columnwidth]{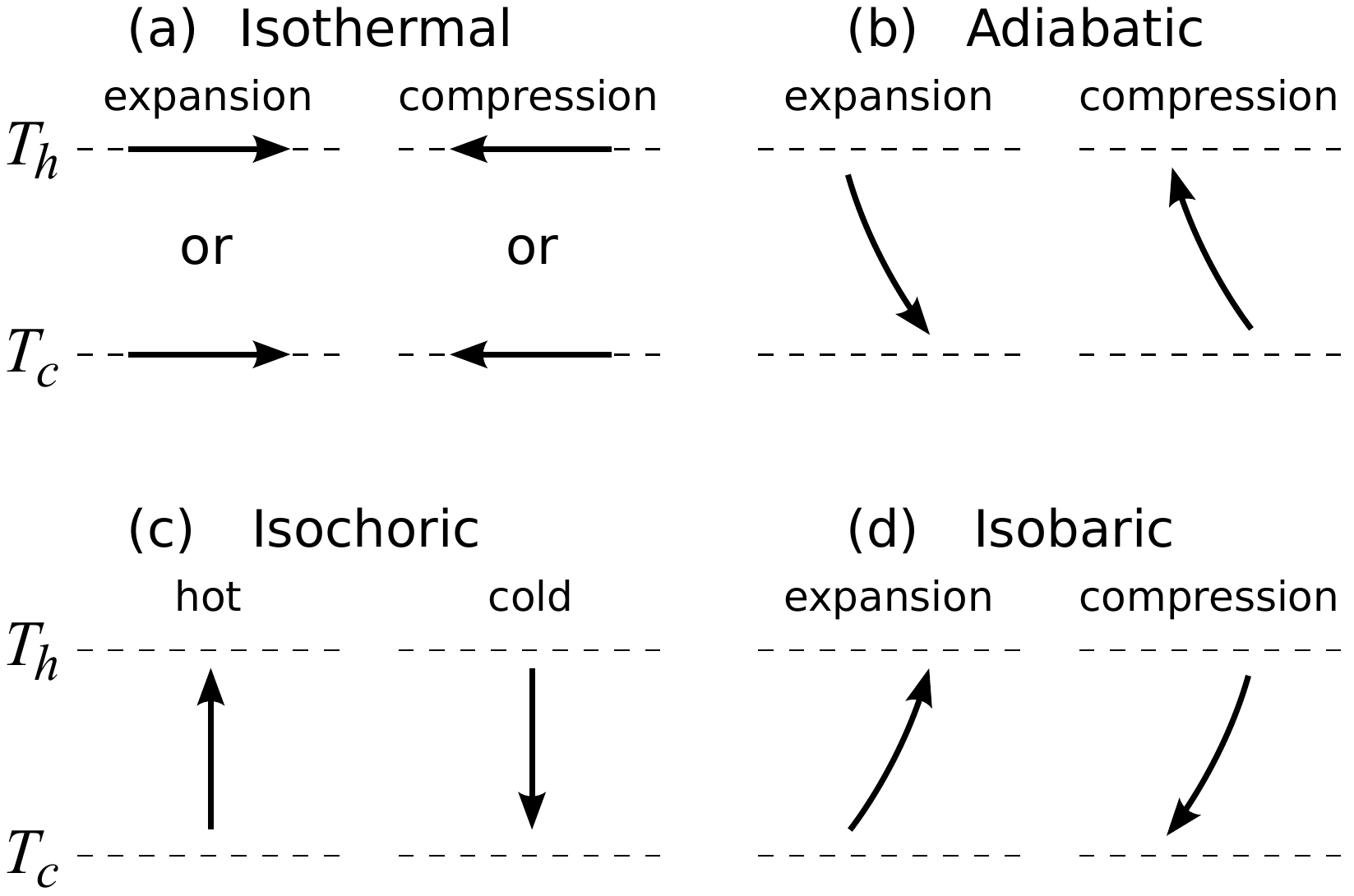}
\caption{\textbf{Considered thermodynamic processes.} Four types of processes [(a) quasistatic isothermal, (b) quasistatic adiabatic, (c) isochoric, and (d) quasistatic isobaric strokes] are shown on the $T$-$\lambda$ plane, where the  $T$-axis is in the vertical direction and $\lambda$-axis is in the horizontal direction. Here, figures show the case in which $\lambda$ is chosen such that increase (decrease) of $\lambda$ corresponds to increase (decrease) of the volume of the system (e.g., the width of a box potential or the inverse of the spring constant of a harmonic oscillator potential is taken as $\lambda$). For other choice of $\lambda$ where increase (decrease) of $\lambda$ corresponds to decrease (increase) of the volume, horizontal axis of the figure should be $\lambda^{-1}$-axis.}
\label{fig:processes}
\end{figure}

\bigskip
\noindent \textbf{Universal bounds on $\eta^{(2)}$ and $\xi^{(2)}$}

Next, we show that the ratios $\eta^{(2)} \equiv \avg{\Delta W^2}/\avg{\Delta Q_{\rm h}^2} $ and $\xi^{(2)} \equiv \avg{\Delta W^2} / \Delta Q^{(2)}$ for the Carnot cycle, given by Eqs.~(\ref{eq:etan}) and (\ref{eq:xi2}), provide upper bounds for these ratios among general quasistatic cycles using the considered working substance. (More precisely, the bound on $\xi^{(2)}$ will be shown for an arbitrary quasistatic cycle and the bound on $\eta^{(2)}$ will be shown for a large class of quasistatic cycles which include most of the typical cycles.) We consider cycles consisting of any of the quasistatic isothermal, quasistatic adiabatic, isochoric, or quasistatic isobaric strokes operating between the temperatures $T_{\rm c}$ and $T_{\rm h}$ (see Fig.~\ref{fig:processes}).

Isobaric processes are those during which the generalized pressure $P \equiv -\avg{\partial H_\lambda/\partial \lambda}$ is constant. Here, the change of $P$ due to the variation of $\lambda$ should be compensated by that due to the heat exchange between the working substance and the heat bath. Since the system is in contact with a heat bath during the isobaric stroke, work fluctuations of the quasistatic isobaric stroke are negligible due to the same reason as the isothermal stroke. As an example, let us choose $\lambda$ such that the increase of $\lambda$ corresponds to the increase of the volume. By expanding the volume (or $\lambda$), the working substance does work. Therefore, the internal energy is used by this expansion, and the pressure of the working substance usually decreases unless there is energy input from the bath. Thus, the heat input should be positive to keep $P$ constant, so that the temperature of the working substance increases. Consequently, a quasistatic isobaric stroke is shown by an upward-sloping curve on the $T$-$\lambda$ plane for such a choice of $\lambda$ [Fig.~\ref{fig:processes}(d)].

To perform quasistatic isobaric processes, we surround the working substance with a ``quasi-adiabatic wall'', which is made of an imperfect heat insulator. When we make a thermal contact between the working substance and the heat bath, heat conduction between them and the thermalization of the working substance occur simultaneously. By surrounding the working substance with the quasi-adiabatic wall and making thermal contact with a heat bath through this wall, we can make the timescale of the heat conduction much larger than that of the thermalization. In this situation, the working substance is thermalized at every moment, i.e., it is always in a canonical state, even if there is an infinitesimally small but continuous heat flux between the working substance and the heat bath that lasts until the temperature of the former reaches that of the latter. Then, quasistatic isobaric processes can be performed by changing $\lambda$ with keeping $P$ constant, which is given by the canonical average of $-\partial H_\lambda/\partial \lambda$ for instantaneous values of the parameter $\lambda$ and the temperature $T$ of the working substance. For example, for a working substance with $I_\lambda(E) = f(\lambda)\, E^\alpha$, the generalized pressure is given by $P = k_{\rm B}T f(\lambda)^{-1} \partial_\lambda f(\lambda)$, where $k_{\rm B}$ is the Boltzmann constant, and the instantaneous values of $\lambda$ and $T$ are related to keep $P$ fixed at the initial value.
In the typical experimental platforms of thermodynamics of small systems, such timescale separation between thermalization in the working substance and heat conduction between the working substance and a bath could be possible for a particle trapped in a harmonic oscillator potential coupled to a thermal environment. The phase space distribution function of this system is always Gaussian provided the initial distribution function is Gaussian. Since the Gaussian phase space distribution function in this system can be identified as a canonical state with some effective temperature $T^{\rm eff}$, the system can be regarded as if it is always in thermal equilibrium at an instantaneous value of $T^{\rm eff}$ at every moment even if $T^{\rm eff}$ is different from the temperature of the bath. On the other hand, the timescale for the heat exchange between the system and bath is determined by the mobility (or diffusion coefficient) of the particle. Thus, by setting the mobility sufficiently small, we could realize the above mentioned timescale separation.

Isochoric processes are those during which $\lambda$ is fixed, and heat is exchanged between the working substance and the heat bath. Since $\lambda$ is fixed through the isochoric stroke, work output and work fluctuations are zero. On the $T$-$\lambda$ plane, an isochoric stroke with the hot (cold) heat bath is shown by an upward (downward) vertical line [Fig.~\ref{fig:processes}(c)]. If we make a direct thermal contact between the working substance and the heat bath, the final state of the isochoric process is a canonical state at the temperature of the bath; however, if we make a thermal contact through the quasi-adiabatic wall, the temperature of the final canonical state can be any value between the initial state and the temperature of the bath.

\begin{table}[t]
  \caption{\textbf{Work and heat fluctuations for each type of process.} Work fluctuation $\avg{\Delta W_{i \rightarrow i+1}^2}$ and heat fluctuation $\avg{\Delta Q_{i \rightarrow i+1}^2}$ for different thermodynamic processes from point $i$ to $i+1$. $\avg{\Delta E_i^2}$ is the variance of the internal energy $E_i$ of the working substance for the canonical distribution at point $i$.}
   \small
   \centering
   \begin{tabular}{lcc}
   \hline\hline
   Process & $\avg{\Delta W_{i \rightarrow i+1}^2}$ & $\avg{\Delta Q_{i \rightarrow i+1}^2}$ \\
   \hline
   (1) Isothermal & $0$ & $\avg{\Delta E_{i}^2} + \avg{\Delta E_{i+1}^2}$ \\
   (2) Adiabatic  & \begin{tabular}{@{}c@{}} $\left[1 - (T_{i+1}/T_i)\right]^2 \avg{\Delta E_i^2}$ \\ $= \left[(T_i/T_{i+1}) - 1\right]^2 \avg{\Delta E_{i+1}^2}$ \end{tabular} & $0$ \\
   (3) Isochoric  & $0$ & $\avg{\Delta E_{i}^2} + \avg{\Delta E_{i+1}^2}$ \\
   (4) Isobaric   & $0$ & $\avg{\Delta E_{i}^2} + \avg{\Delta E_{i+1}^2}$ \\
   \hline
   \end{tabular}
   \label{tab:fluct}
\end{table}

Now we consider the fluctuations of work output and heat input through the above four kinds of processes. Suppose the initial and the final points of the stroke are points $i$ and $i+1$, and the variances of work output and heat input through the stroke $i \rightarrow i+1$ are denoted by $\avg{\Delta W_{i \rightarrow i+1}^2}$ and $\avg{\Delta Q_{i \rightarrow i+1}^2}$, respectively. Regarding the heat exchanging strokes (i.e., quasistatic isothermal, isochoric, and isobaric strokes), since the work fluctuations $\Delta W_{i \rightarrow i+1}$ are negligible and the internal energies $E_i$ and $E_{i+1}$ at their endpoints $i$ and $i+1$ are independent, heat fluctuations of these strokes are given by $\avg{\Delta Q_{i \rightarrow i+1}^2} = \avg{\Delta E_{i+1}^2} + \avg{\Delta E_{i}^2}$. For quasistatic adiabatic strokes, since $W_{i \rightarrow i+1} = E_i - E_{i+1} = [1-(T_{i+1}/T_i)] E_i = [(T_i/T_{i+1})-1] E_{i+1}$ using the AR condition ($E_i/T_i = E_{i+1}/T_{i+1}$) for the third equality, the work fluctuations are given by $\avg{\Delta W_{i \rightarrow i+1}^2} = [1-(T_{i+1}/T_i)]^2 \avg{\Delta E_i^2} = [(T_i/T_{i+1})-1]^2 \avg{\Delta E_{i+1}^2}$. The results of $\avg{\Delta W_{i \rightarrow i+1}^2}$ and $\avg{\Delta Q_{i \rightarrow i+1}^2}$ for each type of process are summarized in Table~\ref{tab:fluct}.

Suppose we have an arbitrary cycle consisting of any sequence of the above four types of processes operating between the bath temperatures $T_{\rm c}$ and $T_{\rm h}$. For the clarity of the discussion, we choose the starting point such that the final stroke is an adiabatic one. In the case of cycles with no adiabatic strokes, trivially $\avg{\Delta W^2}=0$ and thus $\xi^{(2)} = \eta^{(2)}=0$. Among all the $N$ nodes ($k=0,\, 1,\, 2,\, \cdots,\, N$) of the cycle with points $0$ and $N$ being identical, we consider their subsets of both ends of quasistatic adiabatic expansion and compression strokes. For the $i$-th adiabatic expansion (compression) stroke, the initial and the final points are denoted by $J_i$ and $J_i + 1$ ($K_i$ and $K_i + 1$), and the energy and the temperature of these points satisfy the AR condition reading $E_{J_i + 1} = (T_{J_i + 1}/T_{J_i})\, E_{J_i}$\, [$E_{K_i} = (T_{K_i}/T_{K_i + 1})\, E_{K_i + 1}$]. Since only the adiabatic strokes yield work fluctuations, fluctuation of total work output through the cycle is given by
\begin{align}
  \avg{\Delta W^2} =& \sum_i \left[ 1 - (T_{J_i + 1}/T_{J_i}) \right]^2 \avg{\Delta E_{J_i}^2} \nonumber\\
  &+ \sum_j \left[ 1 - (T_{K_j}/T_{K_j + 1}) \right]^2 \avg{\Delta E_{K_j + 1}^2}\,.\label{eq:deltawsq}
\end{align}
Regarding the sum of the heat fluctuations $\Delta Q^{(2)}$, the endpoints of the adiabatic strokes must be connected to the other kinds of strokes, and only these endpoints of the adiabatic strokes contribute to $\Delta Q^{(2)}$. On the other hand, if there are heat-exchanging strokes (those other than quasistatic adiabatic strokes) consecutively, the node connecting them does not contribute to $\Delta Q^{(2)}$ because the internal energies at the final point of the preceding stroke and at the initial point of the following one cancel each other. Thus,
\begin{align}
  \Delta Q^{(2)} =& \sum_i \big( \avg{\Delta E_{J_i}^2} + \avg{\Delta E_{J_i + 1}^2} \big) + \sum_j \big( \avg{\Delta E_{K_j}^2} + \avg{\Delta E_{K_j +  1}^2} \big)\nonumber\\
  =& \sum_i \left[ 1 + (T_{J_i + 1}/T_{J_i})^2 \right] \avg{\Delta E_{J_i}^2} \nonumber\\
  &+ \sum_j \left[ 1 + (T_{K_j}/T_{K_j + 1})^2 \right] \avg{\Delta E_{K_j + 1}^2}\,.\label{eq:deltaqsq}
\end{align}
Since the temperatures $T_{J_i}$ and $T_{J_i + 1}$ ($T_{K_j}$ and $T_{K_j + 1}$) are in the region of $[T_{\rm c}, T_{\rm h}]$ and $T_{J_i} \ge T_{J_i + 1}$ ($T_{K_j + 1} \ge T_{K_j}$), from Eqs.~(\ref{eq:deltawsq}) and (\ref{eq:deltaqsq}) we obtain (see the subsection of ``Derivation of Eqs.~(\ref{eq:deltawsq2}) and (\ref{eq:deltaqsq2})'' in the ``Methods'' section for details)
\begin{align}
  \avg{\Delta W^2} \le \left[ 1 - (T_{\rm c}/T_{\rm h}) \right]^2\, \Big( \sum_i \avg{\Delta E_{J_i}^2} + \sum_j \avg{\Delta E_{K_j + 1}^2} \Big)\,,\label{eq:deltawsq2}\\
  \Delta Q^{(2)} \ge \left[ 1 + (T_{\rm c}/T_{\rm h})^2 \right]\, \Big( \sum_i \avg{\Delta E_{J_i}^2} + \sum_j \avg{\Delta E_{K_j + 1}^2} \Big)\,.\label{eq:deltaqsq2}
\end{align}
Thus, for the ratio $\xi^{(2)} \equiv \avg{\Delta W^2}/\Delta Q^{(2)}$, we finally obtain
\begin{align}
  \xi^{(2)} \le \xi_{\rm C}^{(2)}\,,\label{eq:boundxi}
\end{align}
where $\xi_{\rm C}^{(2)} \equiv [1-(T_{\rm c}/T_{\rm h})]^2/[1+(T_{\rm c}/T_{\rm h})^2]$ is the ratio $\xi^{(2)}$ for the Carnot cycle given by Eq.~(\ref{eq:xi2}). Since a sum of the fluctuations of heat inputs for each stroke trivially satisfies $\sum_{i=0}^{N-1}\avg{\Delta Q_{i \rightarrow i+1}^2} \ge \Delta Q^{(2)}$, another ratio $\tilde{\xi}^{(2)} \equiv \avg{\Delta W^2} / \sum_{i=0}^{N-1}\avg{\Delta Q_{i \rightarrow i+1}^2}$ defined with this quantity is also bounded by $\xi_{\rm C}^{(2)}$: $\tilde{\xi}^{(2)} \le \xi^{(2)} \le \xi_{\rm C}^{(2)}$. Note that $\tilde{\xi}^{(2)} = \xi^{(2)}$ for the Carnot cycle.

Regarding the ratio $\eta^{(2)} \equiv \avg{\Delta W^2}/\avg{\Delta Q_{\rm h}^2}$, the value $\eta_{\rm C}^{(2)}$ for the Carnot cycle gives the maximum value among any cycle in which quasistatic adiabatic expansion (compression) strokes are preceded (followed) by a stroke with the hot bath (if it is not the case, it is possible to have $\eta^{(2)} > \eta_{\rm C}^{(2)}$, see Supplementary Note~1 for an example). This is a natural condition since the thermal energy needs to be taken from (dumped to) a heat bath before the adiabatic expansion (compression). Indeed, most of the typical cycles, including the Otto, Brayton, Stirling, and Ericsson cycles, satisfy this condition. In such cycles, the energy fluctuations at the initial point $J_{i}$ (final point $K_{j} + 1$) of all the quasistatic adiabatic expansion (compression) strokes contribute to $\avg{\Delta Q_{\rm h}^2}$. Thus,
\begin{align}
  \avg{\Delta Q_{\rm h}^2} \ge \sum_i \avg{\Delta E_{J_i}^2} + \sum_j \avg{\Delta E_{K_j + 1}^2}\,.\label{eq:deltaqhsq}
\end{align}
From Eqs.~(\ref{eq:deltawsq2}) and (\ref{eq:deltaqhsq}), we get
\begin{align}
  \eta^{(2)} \le \eta_{\rm C}^{(2)}\,,\label{eq:boundeta}
\end{align}
where $\eta_{\rm C}^{(2)} (= \eta_{\rm C}^2)$ is the ratio $\eta^{(2)}$ for the Carnot cycle given by Eq.~(\ref{eq:etan}) for $n=2$. For the Stirling and Ericsson cycles, $\avg{\Delta W^2} =0$ and thus $\eta^{(2)} = 0$. For the Otto and the Brayton cycles, taking the working substance with $I_\lambda(E)=f(\lambda)\, E^\alpha$ as an example, we get $\eta^{(2)} = [1-(T_2/T_1)]^2$, where $T_1$ and $T_2$ are the initial and the final temperatures of the quasistatic adiabatic expansion stroke. Since $1 \ge T_2/T_1 \ge T_{\rm c}/T_{\rm h}$, this $\eta^{(2)}$ indeed satisfies $\eta^{(2)} \le \eta_{\rm C}^{(2)}$.
Finally, we remark that, besides the Carnot cycle, there also exist other cycles that give the same value of $\eta^{(2)} = \eta_{\rm C}^{(2)}$ and $\xi^{(2)} = \xi_{\rm C}^{(2)}$ (see Supplementary Note~2).

\bigskip
\noindent \textbf{Experimental verification}

Next, we compare our theoretical results against our previous experiment of the Carnot engine using an optically trapped Brownian particle (the so-called Brownian Carnot engine)~\cite{Martinez16}, which is a typical setup for microscopic heat engines. Differences to keep in mind between the Brownian Carnot engine and the true Carnot cycle considered in our work are the following: 1) The working substance of the Brownian engine is always in contact with a thermal environment whose temperature is continuously controllable. 2) The protocol of the Brownian Carnot engine involves two isothermal strokes and two isentropic strokes (i.e., those during which the Shannon entropy of the working substance is constant~\cite{Martinez15}) instead of the isolated adiabatic strokes considered in the present work. Since the working substance of the Brownian heat engine is always in contact with a bath, the variance of work is zero, so that $\eta^{(2)} = 0$ in the quasistatic limit. Nevertheless, there is a correspondence relation between the work distribution of the isolated adiabatic process and the heat distribution of the isentropic strokes for the Brownian Carnot engine in the quasistatic limit (see the subsection of ``Comparison with experimental data'' in the ``Methods'' section for details of the correspondence relation). Therefore, through this correspondence relation, we can simulate the fluctuations of work and heat of the quasistatic Carnot cycle from the experimental data of the Brownian Carnot engine by assuming that the Brownian particle is near equilibrium.

Since the relaxation time ($\sim 0.1$ $\mu$s) of the velocity of the Brownian particle is much shorter than that of the position ($\sim 1$ ms) and the cycle duration in this experiment, we can assume, within the timescale of our interest, that the velocity and the position of the Brownian particle are uncorrelated, and the velocity distribution is always in equilibrium at the instantaneous temperature of the environment. Meanwhile, it is not necessarily the case for the position distribution although it is still Gaussian throughout the whole cycle~\cite{Gardiner_book09}. Therefore, to characterize the state of the working substance, we introduce an effective temperature $T^{\text{eff}}$ defined by the variance of the position $x$ of the Brownian particle as $T^{\text{eff}} = \kappa(t)\avg{x(t)^2}/k_{\rm B}$ (note $\avg{\Delta x^2} = \avg{x^2}$ since $\avg{x} = 0$), where $\kappa(t)$ is the spring constant of the harmonic oscillator trapping potential. Note that $T^{\text{eff}}$ agrees with the temperature of the environment in the quasistatic case.

\begin{figure}[t!]
\centering
\includegraphics[width=0.82\columnwidth]{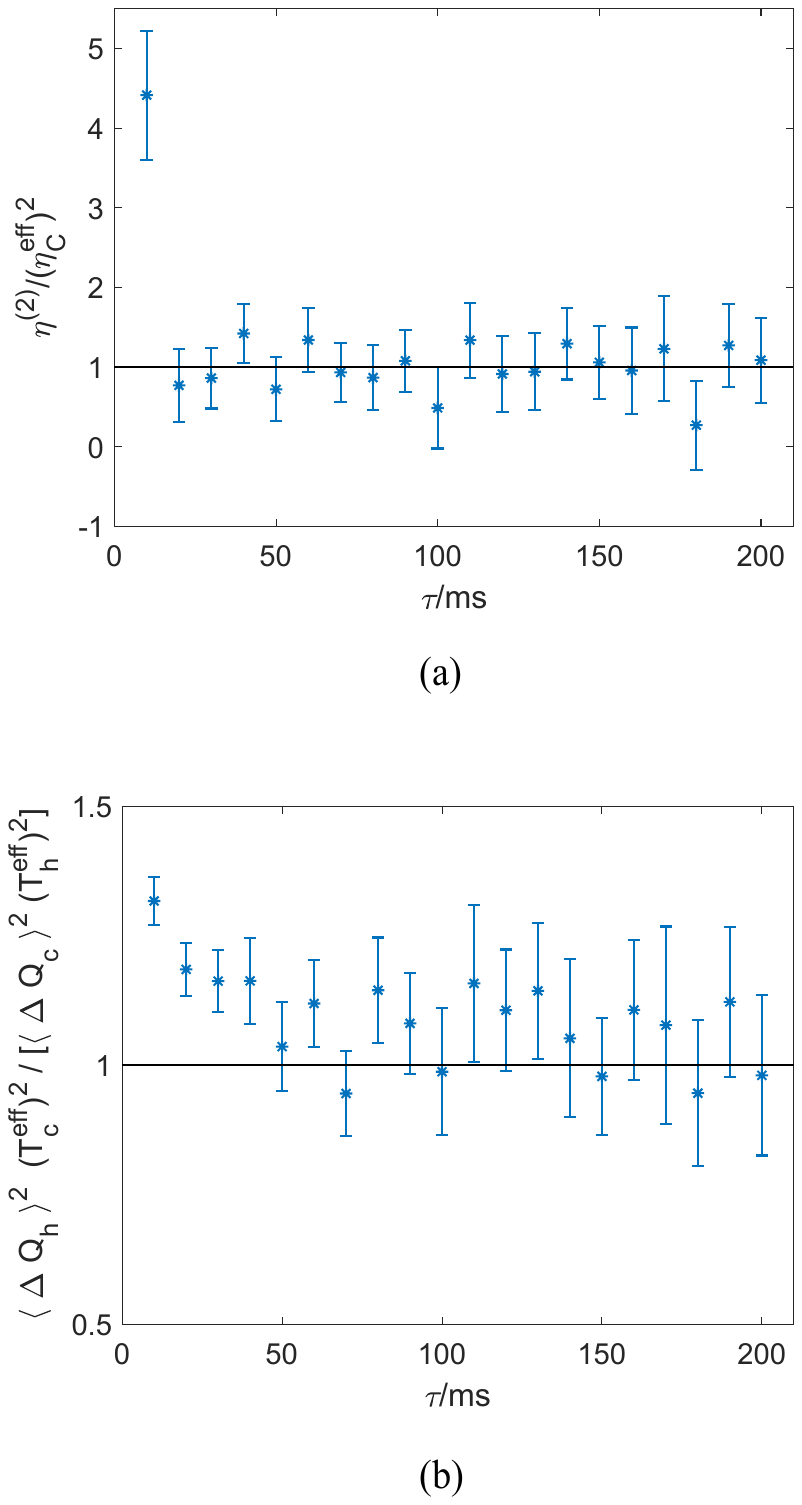}
\caption{\textbf{Experimental verification of the universal relations.} Comparison between the theoretical results and the experimental data for the cycle period ranging from $10$~ms to $200$~ms. 
a) Ratio $\eta^{(2)}/(\eta_{\rm C}^{\text{eff}})^2$ obtained from experimental data of the position of the Brownian particle, which is around unity for large $\tau$ as predicted by Eq.~(\ref{eq:etan}). b) The ratio ${\avg{(\Delta Q_{\rm h}^{\rm exp})^2} (T_{\rm c}^{\text{eff}})^2}/{[\avg{(\Delta Q_{\rm c}^{\rm exp})^2} (T_{\rm h}^{\text{eff}})^2]}$ is around unity for large $\tau$ as predicted by Eq.~(\ref{eq:nthcentralrel}). The error bars in the both panels show the standard error (s.e.m.).
}
\label{fig:experiment}
\end{figure}

Figure~\ref{fig:experiment} shows the comparison between the experimental results and the theoretical predictions of Eqs.~(\ref{eq:etan}) and (\ref{eq:nthcentralrel}) for $n=2$. Figure~\ref{fig:experiment}(a) shows $\eta^{(2)}$ obtained from experimental data for different values of the cycle period $\tau$, normalized by the square of the Carnot efficiency, $(\eta_{\rm C}^\text{eff})^2 \equiv [1 - (T_{\rm c}^{\text{eff}}/T_{\rm h}^{\text{eff}})]^2$, considering the effective temperatures. Further, Fig.~\ref{fig:experiment}(b) shows the ratio between $\avg{(\Delta Q_{\rm c}^{\rm exp})^2}/(T_{\rm c}^{\text{eff}})^2$ and $\avg{(\Delta Q_{\rm h}^{\rm exp})^2}/(T_{\rm h}^{\text{eff}})^2$ for different values of the cycle period $\tau$, where $\avg{(\Delta Q_{\rm c}^{\rm exp})^2}$ and $\avg{(\Delta Q_{\rm h}^{\rm exp})^2}$ represent the variances of heat output and input during the isothermal strokes in the experiment. These figures show that both the ratios, $\eta^{(2)}/(\eta_{\rm C}^\text{eff})^2$ and ${\avg{(\Delta Q_{\rm h}^{\rm exp})^2} (T_{\rm c}^{\text{eff}})^2}/{[\avg{(\Delta Q_{\rm c}^{\rm exp})^2} (T_{\rm h}^{\text{eff}})^2]}$, are around unity, which is consistent with the theoretical prediction except for small $\tau \lesssim 20$~ms for the former and $\tau \lesssim 50$~ms for the latter. It is noted that these ratios are around unity even in the non-quasistatic regime of $100~\mbox{ms} \gtrsim \tau \gtrsim 50~\mbox{ms}$, where the efficiency significantly deviates from $\eta_{\rm C}$ in the experiment (see Fig.~2b in Ref.~\cite{Martinez16}). This result indicates that our theoretical relations (\ref{eq:etan}) and (\ref{eq:nthcentralrel}) would have broader applicability in practice even beyond the quasistatic regime.

\bigskip
\noindent \textbf{\textsf{\large Discussion}}

Statistical characterization of the engine's performance using higher-order moments beyond the mean values is a key issue for understanding the thermodynamics of the mesoscale. We have derived universal relations (\ref{eq:etan}) and (\ref{eq:xi2}) between fluctuations (higher-order central moments) of work and heat in the Carnot cycle. We have also shown that the Carnot cycle provides the universal upper bound for the ratio between the variances of work and heat [Eq.~(\ref{eq:boundxi}) for an arbitrary quasistatic cycles and Eq.~(\ref{eq:boundeta}) for a large class of quasistatic cycles including most of the typical cycles]. This bound might hold even for a wider class of heat engines than those considered in the present work, since the relation similar to Eq.~(\ref{eq:etan}) also holds for steady-state heat engines with nonzero power output \cite{Dechant19,Saryal21}. Furthermore, our experimental test suggests that the results obtained in our work can still hold beyond the quasistatic regime. In addition, our results are applicable to study irreversible non-quasistatic cases through the endo-reversible formalism \cite{Hoffmann08,Xu_eta2}.

Our results provide a guiding principle in the design of reliable and energy-efficient microscopic heat engines, which is an important problem for various topics relevant to the thermodynamics of small systems \cite{Bustamante05,Toyabe15,Martinez17,Fodor21}. From the relations (\ref{eq:etan}) and (\ref{eq:boundeta}), we have $\avg{\Delta W^2}^{1/2} \le \eta_{\rm C}\, \avg{\Delta Q_{\rm h}^2}^{1/2}$. This indicates that we can reduce the fluctuations of the total work output by reducing the fluctuations of heat $\avg{\Delta Q_{\rm h}^2}$ in the heat exchanging stroke(s) with the hot heat bath. For a working substance with the number of states in the form of $I_\lambda(E) = f(\lambda)\, E^\alpha$ as an example, this can be achieved, e.g., by using a trap potential giving a smaller value of $\alpha$: for example, using a box potential ($\alpha=1/2$) instead of a harmonic oscillator one ($\alpha=1$) in one dimension.

Let us illustrate this scheme by taking the Carnot cycle as an example. From Eq.~(\ref{eq:etan}), we have $\avg{\Delta W^n} = \eta_{\rm C}^n \avg{\Delta Q_{\rm h}^n}$. Since the mean value of the heat input, $Q_{\rm h}$ through the quasistatic isothermal stroke depends only on the temperature $T_{\rm h}$ and $\lambda$ but is independent of $\alpha$ as $\avg{Q_{\rm h}} = k_{\rm B}T_{\rm h} \ln{[f(\lambda_1)/f(\lambda_0)]}$ (see the subsection of ``Average and fluctuation of heat input through a quasistatic isothermal stroke for a working substance with $I_\lambda(E)= f(\lambda)E^{\alpha}$'' in the ``Methods'' section), so it is for the mean value of the total work output $\avg{W} = \eta_{\rm C} \avg{Q_{\rm h}}$. On the other hand, the fluctuation of $Q_{\rm h}$ depends only on $T_{\rm h}$ and $\alpha$ as $\avg{\Delta Q_{\rm h}^2} = 2 (k_{\rm B}T_{\rm h})^2 \alpha$ (see the ``Methods'' section). Therefore, by decreasing $\alpha$, we can reduce the fluctuation $\avg{\Delta W^2}$ without reducing the average total work output $W$ and efficiency. In addition to the variance, higher moments $\avg{\Delta W^n}$ with $n>2$ which describe, e.g., the skewness ($n=3$) and the kurtosis ($n=4$) from the Gaussian distribution, can also be reduced at the same time in the above way since the higher moments of $\avg{\Delta Q_{\rm h}^n}$ with $n>2$ also depends only on $T_{\rm h}$ and $\alpha$ [e.g., $\avg{\Delta Q_{\rm h}^3} = 4 (k_{\rm B}T_{\rm h})^3 \alpha$, $\avg{\Delta Q_{\rm h}^4} = 6 (k_{\rm B}T_{\rm h})^4 \alpha (\alpha + 2)$, etc.]. Therefore, the above scheme is beneficial not only for the variance but also for all the moments of $\Delta W$.

As the probability distribution function (pdf) of $W$ for microscopic heat engines significantly deviates from the Gaussian (see, e.g., Fig.~S1 of Ref.~\cite{Martinez16}; note that, while the pdf of the phase space variables is Gaussian, it is not the case for $W$), several higher order moments of $\avg{\Delta W^n}$ for $n=3$ and $n=4$ are also important to characterize the performance of the heat engines. The relation (\ref{eq:etan}) can potentially provide another guiding principle to deal with these higher moments. Finding a scheme to control these higher moments separately from the variance would be an interesting future issue.

\bigskip
\noindent \textbf{\textsf{\large Methods}}

\noindent \textbf{Fluctuations of work and heat in the quasistatic isothermal process}

It has been discussed that the fluctuation of work output throughout the quasistatic isothermal process vanishes as $\sim O(\tau^{-1/2})$ for a long duration $\tau$ of the process \cite{Sekimotobook10,Sekimotobookjp,Holubec18}. Namely, in the limit of large $\tau$, each sample path gives the same value of the work for a given protocol in the quasistatic isothermal process. This is because there is no long-time correlation in the variation of the force exerted by the working substance: this force varies due to thermal fluctuation caused by the contact with a heat bath, and thus the variation does not have a long-time correlation. In the following, we shall show this vanishing fluctuation using the path integral representation.

Let us consider a trajectory $\gamma$ in the phase space between the initial time $t_{\rm init}$ and the final time $t_{\rm fin}$ under the driving of an external control parameter $\lambda(t)$ (whose initial and the final values are $\lambda_{\rm init}$ and $\lambda_{\rm fin}$, respectively). The work output $W$ from the system along this trajectory is given by
\begin{align}
  W[\gamma] \equiv - \int_{t_{\rm init}}^{t_{\rm fin}} dt\, \frac{d\lambda}{dt} \frac{\partial H_\lambda[\gamma]}{\partial\lambda}\,.\label{eq:w_suppl}
\end{align}
The first and the second moments of the work averaged over sample paths are
\begin{align}
  \avg{W} &= \int \mathcal{D}\Gamma\, \mathcal{P}[\gamma]\, W[\gamma]\,,\label{eq:wavr_isoth}\\
  \avg{W^2} &= \int \mathcal{D}\Gamma\, \mathcal{P}[\gamma]\, \left(W[\gamma]\right)^2\,,\label{eq:wsqavr_isoth}
\end{align}
where $\mathcal{P}[\gamma]$ is the probability (density) of the trajectory $\gamma$, and $\int \mathcal{D}\Gamma$ denotes the functional integral with respect to the trajectories.

We set the duration of the process as $\tau \equiv t_{\rm fin} - t_{\rm init} = M \tau_{\rm unit}$ with $M$ being an integer and $\tau_{\rm unit}$ being a sufficiently long time so that the variation of the parameter $\lambda$ is slow enough. (As a prerequisite, $\tau_{\rm unit}$ is taken to be much larger than the correlation time $\tau_{\rm corr}$ of the thermal fluctuation of the force.) Then we shall see the variance $\avg{\Delta W^2}$ of work $W$ vanishes for large $M$. To evaluate the right-hand side of Eqs.~(\ref{eq:wavr_isoth}) and (\ref{eq:wsqavr_isoth}), we discretize the time $\tau_{\rm unit}$ into a sufficiently large number of $N+1$ points, $\{ t_0\equiv t_{\rm init}$, $t_1$, $t_2$, $\cdots$, $t_N \equiv t_{\rm init} + \tau_{\rm unit} \}$, by the time step $\Delta t\equiv t_{n+1} - t_n = \tau_{\rm unit}/N$, which is taken to be much larger than the correlation time $\tau_{\rm corr}$ so that the force exerted by the working substance at different time slices $t_n$ is uncorrelated. Therefore, the probability density $\mathcal{P}[\gamma]$ of the trajectory can be written as a product of the phase space distribution function $P_{\lambda_n}(\Gamma_n;\, t_n)$ at each time slice $t_n$.
The phase space point and the external parameter at time $t_n$ $(0 \le n \le MN)$ are denoted by $\Gamma_n \equiv (\ve{q}_n,\, \ve{p}_n)$ and $\lambda_n$, respectively. Here, $\ve{q}$ and $\ve{p}$ represent the $D$ generalized coordinates and momenta, respectively, in the $2D$-dimensional phase space for the system with $D$ degrees of freedom. Then $\mathcal{D}\Gamma$ reduces to $\mathcal{D}\Gamma \rightarrow \prod_{n=0}^{MN} d\Gamma_n$, where $d\Gamma_n \equiv C d\ve{q}_n\, d\ve{p}_n$ is the phase space volume element at time $t_n$ including the numerical factor $C$ coming from the phase space volume of a microstate.

Thus, the average of work $\avg{W}$ given by Eq.~(\ref{eq:wavr_isoth}) reads
\begin{align}
  \avg{W} \simeq&\, \int \prod_{n=0}^{MN} d\Gamma_n\,\, P_{\lambda_n}(\Gamma_n;\, t_n)\, \left[-\sum_{m=0}^{MN-1}\, \frac{\Delta\lambda_{m/M}}{M}\, \frac{\partial H_{\lambda_m}(\Gamma_m)}{\partial\lambda}\right]\,\nonumber\\
  =&\, - \sum_{m=0}^{MN-1}\, \frac{\Delta \lambda_{m/M}}{M} \int d\Gamma_m\,\, P_{\lambda_m}(\Gamma_m;\, t_m)\, \frac{\partial H_{\lambda_m}(\Gamma_m)}{\partial\lambda}\,,\label{eq:wavr_isoth_2}
\end{align}
where $\Delta \lambda_i$ with $0 \le i \le N$ sets the protocol of the parameter change in the case of $\tau = \tau_{\rm unit}$ with $M=1$: $\Delta \lambda_i \equiv \lambda_{i+1} - \lambda_i$ if $i= m/M$ is an integer; otherwise, $\Delta \lambda_{m/M}$ is between $\Delta \lambda_{[m/M]}$ and $\Delta \lambda_{[m/M]+1}$, where $[m/M]$ is the integer part of $m/M$. Note that the factor of $1/M$ appears because the speed of the parameter change is slowed down by increasing the total duration $\tau$ of the process by a factor of $M$. Similarly, the second moment (\ref{eq:wsqavr_isoth}) becomes
\begin{align}
  \avg{W^2} \simeq&\, \int \prod_{n=0}^{MN} d\Gamma_n\,\, P_{\lambda_n}(\Gamma_n;\, t_n) \left[ -\sum_{m=0}^{MN-1}\, \frac{\Delta\lambda_{m/M}}{M}\, \frac{\partial H_{\lambda_m}(\Gamma_m)}{\partial\lambda} \right]^2\,\nonumber\\
  =&\, \left[ \sum_{m=0}^{MN-1}\, \frac{\Delta \lambda_{m/M}}{M} \int d\Gamma_m\,\, P_{\lambda_m}(\Gamma_m;\, t_m)\, \frac{\partial H_{\lambda_m}(\Gamma_m)}{\partial\lambda} \right]^2\,\nonumber\\
  &\, - \sum_{m=0}^{MN-1}\, \left(\frac{\Delta \lambda_{m/M}}{M}\right)^2 \left[ \int d\Gamma_m\,\, P_{\lambda_m}(\Gamma_m;\, t_m)\, \frac{\partial H_{\lambda_m}(\Gamma_m)}{\partial\lambda} \right]^2\nonumber\\
  &\, + \sum_{m=0}^{MN-1}\, \left(\frac{\Delta \lambda_{m/M}}{M}\right)^2 \int d\Gamma_m\,\, P_{\lambda_m}(\Gamma_m;\, t_m)\, \left[ \frac{\partial H_{\lambda_m}(\Gamma_m)}{\partial\lambda} \right]^2\,.\label{eq:wsqavr_isoth_2}
\end{align}
Here, the second and the third terms of the right-hand side scales as $\sim 1/M$ since the summation $\sum_{m=0}^{MN-1}$ yields a contribution of a factor of $M$, which is multiplied by the factor of $1/M^2$.

From Eqs.~(\ref{eq:wavr_isoth_2}) and (\ref{eq:wsqavr_isoth_2}), we readily see that the variance of work $\avg{\Delta W^2}$ in the quasistatic isothermal process is given by the second and the third terms of Eq.~(\ref{eq:wsqavr_isoth_2}) both of which scale as $\sim 1/M$ and have an opposite sign with each other. Therefore, the variance $\avg{\Delta W^2}$ vanishes as $\sim \tau^{-1}$ or faster:
\begin{align}
  \avg{\Delta W^2} \equiv \avg{W^2} - \avg{W}^2 = O(\tau^{-1})\,,
\end{align}
or the fluctuation $\Delta W$ vanishes as $\sqrt{\avg{\Delta W^2}} = O(\tau^{-1/2})$. As a consequence, fluctuation of $W$ through the quasistatic isothermal process becomes negligible provided the duration $\tau$ of the process is sufficiently long, and $\avg{\Delta W^2} \rightarrow 0$ in the limit of $\tau \rightarrow \infty$.

Next, we shall also discuss the fluctuation of heat during the quasistatic isothermal process. From the first law of thermodynamics for an individual trajectory, the heat absorbed by the system along the trajectory $\gamma$ is given by
\begin{align}
  Q[\gamma] \equiv H_{\lambda_{\rm fin}}(\Gamma_{\rm fin}) - H_{\lambda_{\rm init}}(\Gamma_{\rm init}) + W[\gamma]\,,
\end{align}
where $\Gamma_{\rm init}$ and $\Gamma_{\rm fin}$ are the initial and the final phase space points of the trajectory $\gamma$. The first and the second moments of the heat averaged over sample paths are
\begin{align}
  \avg{Q} &= \int \mathcal{D}\Gamma\, \mathcal{P}[\gamma]\, Q[\gamma]\,,\label{eq:qavr_isoth}\\
  \avg{Q^2} &= \int \mathcal{D}\Gamma\, \mathcal{P}[\gamma]\, \left(Q[\gamma]\right)^2\,.\label{eq:qsqavr_isoth}
\end{align}

The average of heat (\ref{eq:qavr_isoth}) can be written as
\begin{align}
  \avg{Q} = \avg{H_{\lambda_{\rm fin}}}_{\lambda_{\rm fin},\, t_{\rm fin}} - \avg{H_{\lambda_{\rm init}}}_{\lambda_{\rm init},\, t_{\rm init}} + \avg{W}\,,\label{eq:qavr_isoth_2}
\end{align}
where $\avg{W}$ is given by Eq.~(\ref{eq:wavr_isoth_2}), and $\avg{A}_{\lambda,\, t} \equiv \int d\Gamma\,\, P_{\lambda}(\Gamma;\, t)\, A(\Gamma)\,$ is the average of ``$A$'' at time $t$.

Similarly, the second moment (\ref{eq:qsqavr_isoth}) reads
\begin{align}
  \avg{Q^2} =&\, \avg{H_{\lambda_{\rm fin}}^2}_{\lambda_{\rm fin},\, t_{\rm fin}} - 2 \avg{H_{\lambda_{\rm fin}}}_{\lambda_{\rm fin},\, t_{\rm fin}} \avg{H_{\lambda_{\rm init}}}_{\lambda_{\rm init},\, t_{\rm init}} + \avg{H_{\lambda_{\rm init}}^2}_{\lambda_{\rm init},\, t_{\rm init}} + \avg{W^2}\, \nonumber\\
  &\, - 2 \int \mathcal{D}\Gamma\,\, \mathcal{P}[\gamma]\, \left[ H_{\lambda_{\rm fin}}(\Gamma_{\rm fin}) - H_{\lambda_{\rm init}}(\Gamma_{\rm init}) \right]\, \int_{t_{\rm init}}^{t_{\rm fin}} dt\, \frac{d\lambda}{dt}\, \frac{\partial H_{\lambda}[\gamma]}{\partial\lambda}\,.\label{eq:qsqavr_isoth_2}
\end{align}
Here, the fifth term in the right-hand side can be written as
\begin{align}
  & \int \mathcal{D}\Gamma\,\, \mathcal{P}[\gamma]\, \left[ H_{\lambda_{\rm fin}}(\Gamma_{\rm fin}) - H_{\lambda_{\rm init}}(\Gamma_{\rm init}) \right]\, \int_{t_{\rm init}}^{t_{\rm fin}} dt\, \frac{d\lambda}{dt}\, \frac{\partial H_{\lambda}[\gamma]}{\partial\lambda}\nonumber\\
  \simeq&\, \int \prod_{n=0}^{MN} d\Gamma_n\,\, P_{\lambda_n}(\Gamma_n;\, t_n)\, \left[ H_{\lambda_{\rm fin}}(\Gamma_{\rm fin}) - H_{\lambda_{\rm init}}(\Gamma_{\rm init}) \right]\, \sum_{m=0}^{MN-1} \frac{\Delta\lambda_{m/M}}{M} \frac{\partial H_{\lambda_m}(\Gamma_m)}{\partial\lambda}\nonumber\\
  =&\, - \left( \avg{H_{\lambda_{\rm fin}}}_{\lambda_{\rm fin},\, t_{\rm fin}} - \avg{H_{\lambda_{\rm init}}}_{\lambda_{\rm init},\, t_{\rm init}} \right) \avg{W}\nonumber\\
  &\, - \frac{\Delta\lambda_{0}}{M} \left( \avg{H_{\lambda_{\rm fin}}}_{\lambda_{\rm fin},\, t_{\rm fin}} - \avg{H_{\lambda_{\rm init}}}_{\lambda_{\rm init},\, t_{\rm init}} \right) \int d\Gamma_{\rm init}\,\, P_{\lambda_{\rm init}}(\Gamma_{\rm init};\, t_{\rm init})\, \frac{\partial H_{\lambda_{\rm init}}(\Gamma_{\rm init})}{\partial\lambda}\nonumber\\
  &\, + \frac{\Delta\lambda_{0}}{M} \int d\Gamma_{\rm init}\,\, P_{\lambda_{\rm init}}(\Gamma_{\rm init};\, t_{\rm init})\, \left( \avg{H_{\lambda_{\rm fin}}}_{\lambda_{\rm fin},\, t_{\rm fin}} - H_{\lambda_{\rm init}}(\Gamma_{\rm init}) \right)\, \frac{\partial H_{\lambda_{\rm init}}(\Gamma_{\rm init})}{\partial\lambda}\,.\label{eq:qsqavr_isoth_3}
\end{align}
Note that, for the variance $\avg{\Delta Q^2} \equiv \avg{Q^2} - \avg{Q}^2$ of heat $Q$, the contribution from the first term in the right-hand side of Eq.~(\ref{eq:qsqavr_isoth_3}) cancels and only those from the second and the third terms remain, which scale as $\sim 1/M$.
Therefore, from Eqs.~(\ref{eq:qavr_isoth_2}), (\ref{eq:qsqavr_isoth_2}), and (\ref{eq:qsqavr_isoth_3}), the variance of heat in the quasistatic isothermal process reads
\begin{align}
  \avg{\Delta Q^2} =&\, \left[ \avg{H_{\lambda_{\rm fin}}^2}_{\lambda_{\rm fin},\, t_{\rm fin}} - \left(\avg{H_{\lambda_{\rm fin}}}_{\lambda_{\rm fin},\, t_{\rm fin}}\right)^2 \right] + \left[ \avg{H_{\lambda_{\rm init}}^2}_{\lambda_{\rm init},\, t_{\rm init}} - \left(\avg{H_{\lambda_{\rm init}}}_{\lambda_{\rm init},\, t_{\rm init}}\right)^2 \right]\,\, + \avg{\Delta W^2}\nonumber\\
  &\, +2 \frac{\Delta\lambda_{0}}{M} \left( \avg{H_{\lambda_{\rm fin}}}_{\lambda_{\rm fin},\, t_{\rm fin}} - \avg{H_{\lambda_{\rm init}}}_{\lambda_{\rm init},\, t_{\rm init}} \right) \int d\Gamma_{\rm init}\,\, P_{\lambda_{\rm init}}(\Gamma_{\rm init};\, t_{\rm init})\, \frac{\partial H_{\lambda_{\rm init}}(\Gamma_{\rm init})}{\partial\lambda}\nonumber\\
  &\, -2 \frac{\Delta\lambda_{0}}{M} \int d\Gamma_{\rm init}\,\, P_{\lambda_{\rm init}}(\Gamma_{\rm init};\, t_{\rm init})\, \left( \avg{H_{\lambda_{\rm fin}}}_{\lambda_{\rm fin},\, t_{\rm fin}} - H_{\lambda_{\rm init}}(\Gamma_{\rm init}) \right)\, \frac{\partial H_{\lambda_{\rm init}}(\Gamma_{\rm init})}{\partial\lambda}\,\nonumber\\
  =&\, \avg{\Delta H_{\lambda_{\rm fin}}^2}_{\lambda_{\rm fin},\, t_{\rm fin}} + \avg{\Delta H_{\lambda_{\rm init}}^2}_{\lambda_{\rm init},\, t_{\rm init}} + O(\tau^{-1})\,,
\end{align}
with $\avg{\Delta H_{\lambda}^2}_{\lambda,\, t} \equiv \avg{H_{\lambda}^2}_{\lambda,\, t} - (\avg{H_{\lambda}}_{\lambda,\, t})^2$. Thus, $\avg{\Delta Q^2} \rightarrow \avg{\Delta H_{\lambda_{\rm fin}}^2}_{\lambda_{\rm fin},\, t_{\rm fin}} + \avg{\Delta H_{\lambda_{\rm init}}^2}_{\lambda_{\rm init},\, t_{\rm init}}$ in the limit of $\tau \rightarrow \infty$.

\bigskip
\noindent \textbf{Derivation of Eqs.~(\ref{eq:etan}), (\ref{eq:nthcentralrel}), and (\ref{eq:xi2})}

\begin{figure}[t!]
\centering
\includegraphics[width=0.39 \columnwidth]{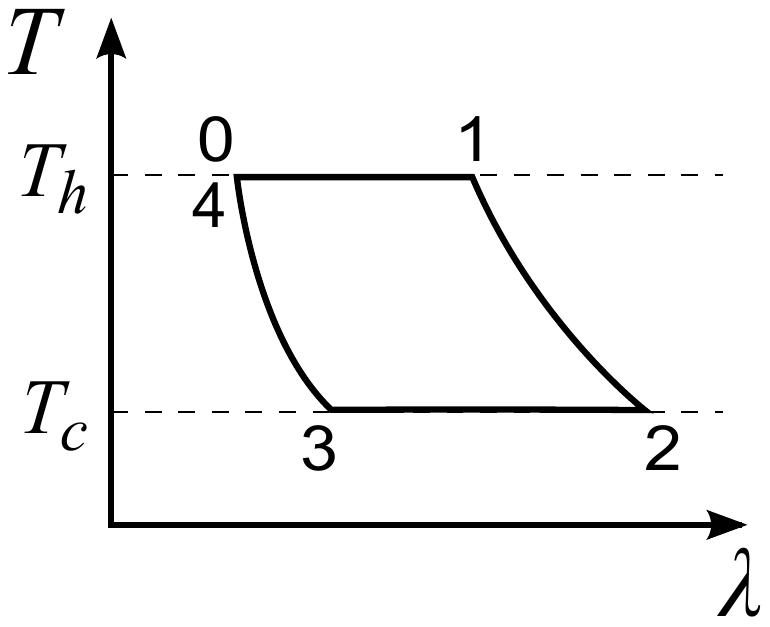}
\caption{\textbf{Carnot cycle.} The temperature versus external control parameter ($T$-$\lambda$) diagram of the Carnot cycle working with a hot heat bath at temperature $T_{\rm h}$ and a cold one at $T_{\rm c}$.
}
\label{fig:carnot_suppl}
\end{figure}

In this subsection, we provide the detailed derivation of the universal relations, Eqs.~(\ref{eq:etan}), (\ref{eq:nthcentralrel}), and (\ref{eq:xi2}), of the central moments of work and heat for the reversible Carnot cycle.
The four strokes of the reversible Carnot cycle (Fig.~\ref{fig:carnot_suppl}) are performed as follows.
$(0)$:~{\it Initial state.---} First, we set the external parameter at $\lambda_0$ and start with a randomly chosen microstate from the canonical ensemble for $H_{\lambda_0}$ at temperature $T_{\rm h}$.
$(0^+ \rightarrow 1^-)$:~{\it Quasistatic isothermal expansion.---} We make a thermal contact between the engine and the hot heat bath with temperature $T_{\rm h}$, then slowly increase the parameter from $\lambda_0$ to $\lambda_1$.
$(1^+ \rightarrow 2^{-})$:~{\it Quasistatic adiabatic expansion.---} We remove the thermal contact between the engine and the bath, and slowly increase the parameter from $\lambda_1$ to $\lambda_2$.
$(2^+ \rightarrow 3^-)$:~{\it Quasistatic isothermal compression.---} At point $2^+$, we make a thermal contact between the engine and the cold heat bath with temperature $T_{\rm c}$, then decrease the parameter from $\lambda_2$ to $\lambda_3$.
$(3^+ \rightarrow 4^-)$:~{\it Quasistatic adiabatic compression.---} We remove the thermal contact between the engine and the bath, and decrease the parameter from $\lambda_3$ to $\lambda_4$. Here, point $4^-$ is equivalent to point $0$: the parameter returns to the initial value, i.e. $\lambda_4=\lambda_0$, and the phase space distribution functions at points $4^-$ and $0$ are the same to close the cycle. Here, we do not include the contribution from point $4^+$ which becomes negligible for a continuous operation over consecutive cycles.

In general, work and heat through each stroke, and the internal energy of the initial and the final states of each stroke are random variables. However, fluctuation $\Delta W_{\rm isoth}$ of the work output $W_{\rm isoth}$ through the quasistatic isothermal process becomes negligible if the duration $\tau$ of the process is sufficiently long \cite{Sekimotobook10,Holubec18} (see also the subsection of ``Fluctuations of work and heat in the quasistatic isothermal process'' in the ``Methods'' section), and it vanishes no slower than $\sim \tau^{-1/2}$, i.e., $\Delta W_{\rm isoth} \equiv W_{\rm isoth} - \avg{W_{\rm isoth}} = O(\tau^{-1/2})$. This is because there is no long-time correlation in the variation of the force exerted by the working substance \cite{Sekimotobook10}. Let us now focus on the quasistatic isothermal expansion stroke $0^+ \rightarrow 1^-$. From the first law of thermodynamics, work output $W_{0 \rightarrow 1}$ by the engine, heat input $Q_{\rm h}$ from the hot heat bath to the working substance, and the internal energy of the working substance $E_0^+$ and $E_1^-$ at the initial and the final state of the stroke should satisfy
\begin{align}
  Q_{\rm h} = E_1^- - E_0^+ + W_{0 \rightarrow 1}\,.\label{eq:1stlaw_hotisoth_suppl}
\end{align}
Since the fluctuation of $W_{0 \rightarrow 1}$ is negligible (i.e., $W_{0 \rightarrow 1} = \avg{W_{0 \rightarrow 1}}$) while $Q_{\rm h}$, $E_0^+$, and $E_1^-$ are not the case, we obtain
\begin{align}
  \Delta Q_{\rm h} \equiv Q_{\rm h} - \avg{Q_{\rm h}} = (E_1^- - \avg{E_1^-}) - (E_0^+ - \avg{E_0^+})\,.\label{eq:deltaqh_suppl}
\end{align}
Here, $\avg{Q_{\rm h}}$ is the ensemble average of $Q_{\rm h}$ over possible sample paths, and $\avg{E_i^{\pm}}$ is the average of $E_i^{\pm}$ over the canonical ensemble at point $i^\pm$. Note that $E_i^+ = E_i^-$ because of the first law of thermodynamics together with the fact that the work for making and removing a thermal contact with a heat bath is negligible \cite{Sekimotobook10}. [However, $\avg{E_i^+}=\avg{E_i^-}$ is sufficient to prove the universal relations (\ref{eq:etan}), (\ref{eq:nthcentralrel}), and (\ref{eq:xi2}). The condition $E_i^+ = E_i^-$ is required only in the proof of the universal bounds given by Eqs.~(\ref{eq:boundxi}) and (\ref{eq:boundeta}).]

Next, we consider the total work output $W$ through the whole cycle:
\begin{align}
  W = W_{0 \rightarrow 1} + W_{1 \rightarrow 2} + W_{2 \rightarrow 3} + W_{3 \rightarrow 4}\,,
\end{align}
where $W_{i \rightarrow i+1}$ is work output through the stroke from point $i^+$ to $(i+1)^-$.
Since the fluctuations of $W_{0 \rightarrow 1}$ and $W_{2 \rightarrow 3}$ by the quasistatic isothermal strokes are negligible, we obtain
\begin{align}
  \Delta W \equiv W - \avg{W} = (W_{1 \rightarrow 2} - \avg{W_{1 \rightarrow 2}}) + (W_{3 \rightarrow 4} - \avg{W_{3 \rightarrow 4}})\,.\label{eq:dw_suppl}
\end{align}
Here, the strokes $1^+ \rightarrow 2^-$ and $3^+ \rightarrow 4^-$ are quasistatic adiabatic processes. Since there is no heat exchange between the working substance and the heat bath during these strokes, $W_{1 \rightarrow 2}$, for example, reads
\begin{align}
  W_{1 \rightarrow 2} = E_1^{+} - E_2^{-}\,.\label{eq:1stlaw_adexp_suppl}
\end{align}
In addition, for our working substance, since the initial and the final state of the quasistatic adiabatic stroke should satisfy the AR condition, we have $E_1^+/T_{\rm h} = E_2^-/T_{\rm c}$. From this relation and Eq.~(\ref{eq:1stlaw_adexp_suppl}), we get
\begin{align}
  W_{1 \rightarrow 2} = \left[ 1 - (T_{\rm c}/T_{\rm h}) \right]\, E_1^+\,.\label{eq:w12_suppl}
\end{align}
Similarly, for the stroke $3^+ \rightarrow 4^-$, we get $W_{3 \rightarrow 4} = E_3^+ - E_4^- = \left[ (T_{\rm c}/T_{\rm h}) - 1 \right]\, E_4^-$. Thus Eq.~(\ref{eq:dw_suppl}) reads
\begin{align}
  \Delta W = \left[ 1 - (T_{\rm c}/T_{\rm h}) \right]\, \left[ (E_1^+ - \avg{E_1^+}) - (E_4^- - \avg{E_4^-}) \right]\,.\label{eq:deltaw_suppl}
\end{align}
From Eqs.~(\ref{eq:deltaqh_suppl}) and (\ref{eq:deltaw_suppl}) together with the fact that $E_0^+$ and $E_4^-$ are statistically equivalent random variables (i.e., random variables following the same probability distribution function), we finally obtain Eq.~(\ref{eq:etan}):
\begin{align}
  \eta^{(n)} \equiv \frac{\avg{(\Delta W)^n}}{\avg{(\Delta Q_{\rm h})^n}} = \left( 1 - \frac{T_{\rm c}}{T_{\rm h}} \right)^n\,.\nonumber
\end{align}
for any integer $n \ge 2$.

For the quasistatic isothermal compression stroke $2^+ \rightarrow 3^-$, heat output $Q_{\rm c}$ from the working substance to the cold heat bath through this stroke is given by
\begin{align}
  Q_{\rm c} = E_2^+ - E_3^- - W_{2 \rightarrow 3}\,.\label{eq:1stlaw_coldisoth_suppl}
\end{align}
Since the fluctuation of $W_{2 \rightarrow 3}$ is negligible, we get
\begin{align}
  \Delta Q_{\rm c} \equiv Q_{\rm c} - \avg{Q_{\rm c}} = (E_2^+ - \avg{E_2^+}) - (E_3^- - \avg{E_3^-})\,.\label{eq:deltaqc_suppl}
\end{align}
The AR condition for the quasistatic adiabatic strokes $1^+ \rightarrow 2^-$ and $3^+ \rightarrow 4^-$ reads $E_1^+/T_{\rm h} = E_2^-/T_{\rm c}$ and $E_3^+/T_{\rm c} = E_4^-/T_{\rm h}$, respectively. Since $E_2^{\pm}$ and $E_3^{\pm}$ are independent and $E_{2, 3}^- = E_{2, 3}^+$, Eq.~(\ref{eq:deltaqc_suppl}) with the above conditions leads to Eq.~(\ref{eq:nthcentralrel}):
\begin{align}
  \avg{(\Delta Q_{\rm c})^n} =& \left(T_{\rm c}/T_{\rm h}\right)^n \avg{\left[ (E_1 - \avg{E_1}) - (E_0 - \avg{E_0}) \right]^n}\,\nonumber\\
  =& \left(T_{\rm c}/T_{\rm h}\right)^n \avg{(\Delta Q_{\rm h})^n}\,,\nonumber
\end{align}
which can be rewritten in the form analogous to the ``central'' relation of thermodynamics \cite{Feynmanlect}, $\avg{Q_{\rm h}}/T_{\rm h} = \avg{Q_{\rm c}}/T_{\rm c}$, as:
\begin{align}
  \frac{\avg{(\Delta Q_{\rm h})^n}}{T_{\rm h}^n} = \frac{\avg{(\Delta Q_{\rm c})^n}}{T_{\rm c}^n}\,.\nonumber
\end{align}
Note that, since the temperatures $T_{\rm c}$ and $T_{\rm h}$ are arbitrary, this relation holds for two arbitrary quasistatic isotherms connected by two quasistatic adiabats.
The sum $\Delta Q^{(2)}$ of the heat fluctuations over the separated sequances of heat exchanging strokes is given by [same as Eq.~(\ref{eq:deltaq2})]
\begin{align}
    \Delta Q^{(2)} \equiv& \avg{\Delta Q_{\rm h}^2} + \avg{\Delta Q_{\rm c}^2}
  = \left[ 1 + (T_{\rm c}/T_{\rm h})^2 \right] \avg{\Delta Q_{\rm h}^2}\,.\nonumber
\end{align}
From this equation and Eq.~(\ref{eq:etan}), we obtain Eq.~(\ref{eq:xi2}):
\begin{align}
  \xi^{(2)} \equiv \frac{\avg{\Delta W^2}}{\Delta Q^{(2)}} = \frac{\left[1- \left(T_{\rm c}/T_{\rm h}\right)\right]^2}{1+ (T_{\rm c}/T_{\rm h})^2 }\,.
\end{align}

\bigskip
\noindent \textbf{Derivation of Eqs.~(\ref{eq:deltawsq2}) and (\ref{eq:deltaqsq2})}

Here we show the detailed derivation of Eqs.~(\ref{eq:deltawsq2}) and (\ref{eq:deltaqsq2}). Since the temperatures $T_{J_i}$ and $T_{J_i + 1}$ ($T_{K_j}$ and $T_{K_j + 1}$) are in the region of $[T_{\rm c}, T_{\rm h}]$ and $T_{J_i} \ge T_{J_i + 1}$ ($T_{K_j + 1} \ge T_{K_j}$), we get $1 \ge T_{J_i + 1}/T_{J_i} \ge T_{\rm c}/T_{J_i} \ge T_{\rm c}/T_{\rm h}$ ($1 \ge T_{K_j}/T_{K_j + 1} \ge T_{\rm c}/T_{\rm h}$). Therefore,
\begin{align}
  [1-(T_{J_i + 1}/T_{J_i})]^2,\,\,\, [1-(T_{K_j}/T_{K_j + 1})]^2\, \le\, [1 - (T_{\rm c}/T_{\rm h})]^2\,,
\end{align}
and
\begin{align}
  1+(T_{J_i + 1}/T_{J_i})^2,\,\,\, 1+(T_{K_j}/T_{K_j + 1})^2\, \ge\, 1 + (T_{\rm c}/T_{\rm h})^2\,.
\end{align}
Applying these relations to Eqs.~(\ref{eq:deltawsq}) and (\ref{eq:deltaqsq}):
\begin{align}
  \avg{\Delta W^2} =& \sum_i \left[ 1 - (T_{J_i + 1}/T_{J_i}) \right]^2 \avg{\Delta E_{J_i}^2} + \sum_j \left[ 1 - (T_{K_j}/T_{K_j + 1}) \right]^2 \avg{\Delta E_{K_j + 1}^2}\,,\nonumber\\
  \Delta Q^{(2)} =& \sum_i \left[ 1 + (T_{J_i + 1}/T_{J_i})^2 \right] \avg{\Delta E_{J_i}^2} + \sum_j \left[ 1 + (T_{K_j}/T_{K_j + 1})^2 \right] \avg{\Delta E_{K_j + 1}^2}\,,\nonumber
\end{align}
we finally obtain Eqs.~(\ref{eq:deltawsq2}) and (\ref{eq:deltaqsq2}):
\begin{align}
  \avg{\Delta W^2} \le& \left[ 1 - (T_{\rm c}/T_{\rm h}) \right]^2\, \Big( \sum_i \avg{\Delta E_{J_i}^2} + \sum_j \avg{\Delta E_{K_j + 1}^2} \Big)\,,\nonumber\\
  \Delta Q^{(2)} \ge& \left[ 1 + (T_{\rm c}/T_{\rm h})^2 \right]\, \Big( \sum_i \avg{\Delta E_{J_i}^2} + \sum_j \avg{\Delta E_{K_j + 1}^2} \Big)\,.\nonumber
\end{align}

\bigskip
\noindent \textbf{Average and fluctuation of heat input through a quasistatic isothermal stroke for a working substance with $I_\lambda(E)= f(\lambda)E^{\alpha}$}

In this subsection, we derive the average and the fluctuation of heat input $Q_{\rm isoth}$ in the quasistatic isothermal processes for a working substance with the number of states $I_\lambda(E) = f(\lambda) E^\alpha$. Here, $f$ is a function of the external parameter $\lambda$, and $\alpha$ is a real constant, which are fixed for each setup of the system. Namely, the form of $f$ and the value of $\alpha$ are determined by, e.g., the shape of the trapping potential. For example, $f(d)=\sqrt{8m}\, d$ and $\alpha=1/2$ for a particle (mass $m$) in a 1-dimensional box potential with width $d$, and $f(\omega^{-1})=2\pi\omega^{-1}$ and $\alpha=1$ for a particle in a 1-dimensional harmonic oscillator potential with frequency $\omega$.

For the working substance with $I_\lambda(E) = f(\lambda) E^\alpha$, the density of state $g_\lambda(E)$ reads
\begin{align}
  g_\lambda(E) \equiv \frac{\partial I_\lambda(E)}{\partial E} = \int d\Gamma\,\, \delta\left(E - H_{\lambda}(\Gamma)\right) = \alpha\, f(\lambda)\, E^{\alpha - 1}\,.\label{eq:g_suppl}
\end{align}
Then, the canonical distribution $P_{\beta,\, \lambda}^{\rm eq}$ at the inverse temperature $\beta \equiv 1/k_{\rm B}T$ and the external parameter $\lambda$ is
\begin{align}
  P_{\beta,\, \lambda}^{\rm eq}(E) = \frac{g_\lambda(E)}{Z_{\beta,\, \lambda}} \mathrm{e}^{-\beta E} = \frac{\beta^\alpha\, E^{\alpha - 1}}{\Gamma(\alpha)} \mathrm{e}^{-\beta E}\,,\label{eq:canonical_suppl}
\end{align}
with $Z_{\beta,\, \lambda}$ being the partition function given by
\begin{align}
  Z_{\beta,\, \lambda} = \int_0^\infty dE\,\, g_\lambda(E)\, \mathrm{e}^{-\beta E} = \alpha\, \Gamma(\alpha)\, f(\lambda)\, \beta^{-\alpha}\,,\label{eq:partition_suppl}
\end{align}
where $\Gamma(\alpha) \equiv \int_0^\infty dx\, x^{\alpha - 1} \mathrm{e}^{-x}$ is the gamma function.

For the canonical distribution given by Eq.~(\ref{eq:canonical_suppl}), the average of the internal energy $E_i$ and of its square $E_i^2$ at point $i$ with temperature $T_i$ and parameter $\lambda_i$ can be calculated as
\begin{align}
  \avg{E_i} &= \int_0^\infty dE\,\, E\, P_{\beta_i,\, \lambda_i}^{\rm eq}(E) = k_{\rm B}T_i \alpha\,,\label{eq:ei_suppl}\\
  \avg{E_i^2} &= \int_0^\infty dE\,\, E^2\, P_{\beta_i,\, \lambda_i}^{\rm eq}(E) = (k_{\rm B}T_i)^2\, (\alpha+1)\, \alpha\,.
\end{align}
Therefore, the variance $\avg{\Delta E_i^2}$ of $E_i$ reads
\begin{align}
  \avg{\Delta E_i^2} = \avg{E_i^2} - \avg{E_i}^2 = (k_{\rm B}T_i)^2 \alpha\,.\label{eq:deltaesq_suppl}
\end{align}

Suppose the initial and the final nodes of the quasistatic isothermal stroke are denoted by points $1$ and $2$, the parameter $\lambda$ at points $1$ and $2$ are denoted by $\lambda_1$ and $\lambda_2$ respectively, and the temperature of the working substance is constant at $T$. For the working substance considered, the average of the internal energy $E_i$ of the working substance in an equilibrium state at point $i$ in general is given by Eq.~(\ref{eq:ei_suppl}): $\avg{E_i} = k_{\rm B}T_i\alpha$. Therefore, for quasistatic isothermal strokes, the change in the average of the internal energy through the stroke is zero:
\begin{align}
  \avg{E_2} - \avg{E_1} = 0\,.
\end{align}
Thus, from the first law of thermodynamics, the average of the heat input $Q_{\rm isoth}$ is equal to the average of the work output $W_{\rm isoth}$ through the quasistatic isothermal stroke:
\begin{align}
  \avg{Q_{\rm isoth}} = \avg{W_{\rm isoth}}\,.\label{eq:1stlaw_qsisoth_suppl}
\end{align}

For quasistatic isothermal processes, the work output given by Eqs.~(\ref{eq:w_suppl}) and (\ref{eq:wavr_isoth}) reads
\begin{align}
  \avg{W_{\rm isoth}} &= -\int_{\lambda_1}^{\lambda_2} d\lambda \int d\Gamma\,\, \frac{\partial H_\lambda}{\partial \lambda}\, P_{\beta,\, \lambda}^{\rm eq}(\Gamma)\nonumber\\
  &= \int_{\lambda_1}^{\lambda_2} d\lambda\,\, \frac{1}{Z_{\beta,\, \lambda}} \frac{1}{\beta} \frac{\partial Z_{\beta,\, \lambda}}{\partial \lambda}\nonumber\\
  &= k_{\rm B}T\, \ln{\frac{f(\lambda_2)}{f(\lambda_1)}}\,,\label{eq:wavr_qsisoth_suppl}
\end{align}
where
\begin{align}
  P_{\beta,\, \lambda}^{\rm eq}(\Gamma) \equiv \frac{\mathrm{e}^{-\beta H_\lambda(\Gamma)}}{Z_{\beta,\, \lambda}}
\end{align}
is the canonical distribution with respect to the phase space point $\Gamma$. From Eqs.~(\ref{eq:1stlaw_qsisoth_suppl}) and (\ref{eq:wavr_qsisoth_suppl}), we finally get
\begin{align}
  \avg{Q_{\rm isoth}} = k_{\rm B}T \ln{\frac{f(\lambda_2)}{f(\lambda_1)}}\,.\label{eq:qavr_qsisoth_suppl}
\end{align}
The same result can also be obtained from the change in the entropy,
\begin{align}
  \avg{S_i} \equiv -k_{\rm B} \int d\Gamma\,\, P_{\beta,\, \lambda_i}^{\rm eq}(\Gamma)\, \ln{P_{\beta,\, \lambda_i}^{\rm eq}(\Gamma)} = k_{\rm B} (\alpha + \ln{Z_{\beta,\, \lambda}})\,,
\end{align}
between the initial and the final states of the stroke:
\begin{align}
  \avg{Q_{\rm isoth}} = T (\avg{S_2} - \avg{S_1})\,.
\end{align}

Next, we derive the variance $\avg{\Delta Q_{\rm isoth}^2}$ of heat input through the quasistatic isothermal stroke for the working substance considered. According to Table~\ref{tab:fluct}, $\avg{\Delta Q_{\rm isoth}^2}$ is given by a sum of the variances of the internal energy at the initial and the final states of the stroke. From Eq.~(\ref{eq:deltaesq_suppl}) with $T \equiv T_1 = T_2$, we obtain
\begin{align}
  \avg{\Delta Q_{\rm isoth}^2} = \avg{\Delta E_1^2} + \avg{\Delta E_2^2} = 2 (k_{\rm B} T)^2 \alpha\,.\label{eq:deltaqsq_qsisoth_suppl}
\end{align}

From Eqs.~(\ref{eq:qavr_qsisoth_suppl}) and (\ref{eq:deltaqsq_qsisoth_suppl}), we can see that the mean value $\avg{Q_{\rm isoth}}$ of the heat input $Q_{\rm isoth}$ through the quasistatic isothermal stroke depends only on $T$ and $\lambda$ while its variance $\avg{\Delta Q_{\rm isoth}^2}$ depends only on $T$ and $\alpha$.

\bigskip
\noindent \textbf{Integral fluctuation theorems and $\eta_{\rm C}^{(n)}$}

Following the treatment in Ref.~\cite{Jarzynski00}, the Jarzynski equality can be extended to the system coupled with two heat baths. Assuming that the coupling Hamiltonians between the system and the baths are negligible (the so-called weak-coupling case), the resulting integral fluctuation theorem reads \cite{Pal17} (see also Refs.~\cite{Sinitsyn11,Campisi14})
\begin{align}
  \avg{\mathrm{e}^{-\Sigma}} = 1\,\label{eq:integralFT_suppl}
\end{align}
with the stochastic entropy production $\Sigma$ including the total entropy change in the system and the baths through the whole process whose specific form will be defined below. Note that Eq.~(\ref{eq:integralFT_suppl}) is applicable to the protocol where the cycle starts with a canonical state at some temperature and the system is equilibrated again at the same temperature as the initial one at the end of the whole process.

For reversible processes as the Carnot cycle considered in the present work which satisfies the AR condition, it is known that the mean value of the entropy production is zero:
\begin{align}
  \avg{\Sigma} = 0\,.\label{eq:zeromeanEP_suppl}
\end{align}
From Jensen's inequality together with Eq.~(\ref{eq:zeromeanEP_suppl}), the lhs of Eq.~(\ref{eq:integralFT_suppl}) leads to
\begin{align}
  \avg{\mathrm{e}^{-\Sigma}} \ge \mathrm{e}^{-\avg{\Sigma}} = 1\,.
\end{align}
Therefore, for reversible processes we obtain
\begin{align}
  \avg{\mathrm{e}^{-\Sigma}} = \mathrm{e}^{-\avg{\Sigma}}\,\label{eq:integralFT2_suppl}
\end{align}
with $\avg{\Sigma} = 0$. One can show that Eqs.~(\ref{eq:zeromeanEP_suppl}) and (\ref{eq:integralFT2_suppl}) simultaneously hold if and only if the stochastic entropy production $\Sigma$ is deterministically zero:
\begin{align}
  \Sigma =0\label{eq:zerosigma_suppl}
\end{align}
for any sample path, i.e., the distribution function $P_{\Sigma}(\sigma)$ of $\Sigma$ is
\begin{align}
  P_{\Sigma}(\sigma) = \delta(\sigma)\,,
\end{align}
where $\delta$ is the Dirac delta function.

For a cyclic protocol starting from the canonical state at $T_{\rm c}$ (protocol 1), $\Sigma$ in Eq.~(\ref{eq:integralFT_suppl}) is given by
\begin{align}
  \Sigma = \frac{1}{T_{\rm c}} \left[\left(1 - \frac{T_{\rm c}}{T_{\rm h}} \right) Q_{\rm h}^{[1]} - W^{[1]} \right]\,.\label{eq:sigma1_suppl}
\end{align}
On the other hand, $\Sigma$ for a protocol starting from the canonical state at $T_{\rm h}$ (protocol 2) is
\begin{align}
  \Sigma = \frac{1}{T_{\rm h}} \left[\left(\frac{T_{\rm h}}{T_{\rm c}} - 1 \right) Q_{\rm c}^{[2]} - W^{[2]} \right]\,.\label{eq:sigma2_suppl}
\end{align}
Here, the superscripts ``$[i]$'' ($i=1$ and $2$) in Eqs.~(\ref{eq:sigma1_suppl}) and (\ref{eq:sigma2_suppl}) mean the random variables associated to protocol $i$. As has been pointed out below Eq.~(\ref{eq:xi2}), it is noted that the random variables in Eqs.~(\ref{eq:sigma1_suppl}) and (\ref{eq:sigma2_suppl}) for different initial points (e.g., $W^{[1]}$ and $W^{[2]}$) should be regarded as different random variables, which belong to different ensembles. This is clear from the fact that the fluctuation $\Delta E$ of the internal energy change through one cycle, $E = (Q_{\rm h} - Q_{\rm c}) - W$, depends on the temperature of the initial point, so that work and heat for different protocols with different initial points are different random variables, which could have different amount of fluctuations. Unlike the mean values which are constrained by $\avg{E} = 0$ for a cycle, there is no such a constraint on the fluctuations. Therefore, in the discussion of the fluctuations for a single cycle, the effects of the end points of the process are important.

From Eq.~(\ref{eq:zerosigma_suppl}) and (\ref{eq:sigma1_suppl}), we obtain
\begin{align}
  W^{[1]} = \left( 1 - \frac{T_{\rm c}}{T_{\rm h}} \right) Q_{\rm h}^{[1]}\,,\label{eq:w1qh1_suppl}
\end{align}
and similarly from Eq.~(\ref{eq:zerosigma_suppl}) and (\ref{eq:sigma2_suppl}), we obtain
\begin{align}
  W^{[2]} = \left( \frac{T_{\rm h}}{T_{\rm c}} - 1 \right) Q_{\rm c}^{[2]}\,.\label{eq:w2qc2_suppl}
\end{align}
Thus, from Eq.~(\ref{eq:w1qh1_suppl}), we can obtain the universal relation of $\eta^{(n)}$ given by Eq.~(\ref{eq:etan}) for protocol 1:
\begin{align}
  \eta^{(n)}(\mbox{for protocol 1}) = \frac{\avg{ (\Delta W^{[1]})^n }_1}{\avg{ (\Delta Q_{\rm h}^{[1]})^n }_1} = \left( 1 - \frac{T_{\rm c}}{T_{\rm h}} \right)^n\,,
\end{align}
where $\avg{\cdots}_i$ represents the ensemble average of protocol $i$.
However, it is not trivial to derive the same relation for different protocols with an initial state other than the canonical state at $T_{\rm c}$ from the integral fluctuation theorem.

Regarding the other universal relations given by Eqs.~(\ref{eq:nthcentralrel}) and (\ref{eq:xi2}), it is also non-trivial to derive these relations from the integral fluctuation theorem since $W^{[1]}$ and $W^{[2]}$ in Eqs.~(\ref{eq:w1qh1_suppl}) and (\ref{eq:w2qc2_suppl}) are different random variables so that $\avg{(\Delta W^{[1]})^n}_1$ and $\avg{(\Delta W^{[2]})^n}_2$ cannot be identified trivially.

Unlike the above derivation based on the integral fluctuation theorem, all the random variables of $W$, $Q_{\rm h}$, and $Q_{\rm c}$ for the same protocol can be handled at the same time in the bottom-up approach employed in the subsection of ``Derivation of Eqs.~(\ref{eq:etan}), (\ref{eq:nthcentralrel}), and (\ref{eq:xi2})'' of the ``Methods'' section. In addition, with this bottom-up approach, deriving the relations (\ref{eq:etan}), (\ref{eq:nthcentralrel}), and (\ref{eq:xi2}) for any other initial point is straightforward.

\bigskip
\noindent \textbf{Comparison with experimental data}

\noindent \textit{Correspondence relations.}

We provide detailed derivation of the correspondence relations between the work distribution of the true adiabatic process in the Carnot cycle and the heat distribution of the isentropic process in the Brownian Carnot cycle (hereafter, Brownian isentropic process for short) \cite{Martinez16} in the quasistatic limit. In both the true adiabatic process and the Brownian isentropic process, the Shannon entropy is constant and the system is in canonical state throughout the process in the quasistatic limit. Therefore, for these processes starting at the same temperature $T$ and the external parameter $\lambda$, their paths on the $T$-$\lambda$ plane are the same.

Regarding the true adiabatic process, since the working substance is isolated during the process, its work output $W^{\text{true}}_{\text{ad}}$ is given by
\begin{equation}
	W^{\text{true}}_{\text{ad}} = E_{\text{init}} - E_{\text{fin}} = \left( 1 - \frac{T_{\text{fin}}}{T_{\text{init}}} \right) E_{\text{init}},
\end{equation}
where $E_\text{init}$ ($E_\text{fin}$) and $T_{\text{init}}$ ($T_{\text{fin}}$) are the energy and temperature of the initial (final) state of the true adiabatic process, respectively. Here, we have used the AR condition, $E_{\text{init}}/T_{\text{init}} = E_{\text{fin}}/T_{\text{fin}}$, for the second equality.
Since the probability distribution of $E_{\text{init}}$ is the canonical state with the inverse temperature $\beta_{\text{init}}=(k_{\rm B} T_{\text{init}})^{-1}$, i.e. $P(E_{\text{init}}) \propto \mathrm{e}^{-\beta_{\text{init}} E_{\text{init}}}$ ($E_{\text{init}}>0$), the work distribution function $P_{W^{\text{true}}_{\text{ad}} }(w)$ reads
\begin{equation}
	P_{W^{\text{true}}_{\text{ad}} }(w) = \frac{1}{k_{\rm B} | T_{\text{fin}}-T_{\text{init}} |} \mathrm{e}^{-\frac{1}{k_{\rm B} (T_{\text{init}}-T_{\text{fin}}) } w },
\end{equation}
for $w > 0$ [$w \le 0$] and $P_{W^{\text{true}}_{\text{ad}} }(w) = 0$ for $w \le 0$ [$w > 0$] when $T_{\text{init}} > T_{\text{fin}}$ (quasistatic adiabatic expansion) [$T_{\text{init}} \le T_{\text{fin}}$ (quasistatic adiabatic compression)].

On the other hand, in the Brownian isentropic process, which is an experimental counterpart of the true adiabatic process \cite{Martinez15}, the working substance is always in contact with a heat bath with continuously variable temperature. In this process, the temperature and the external parameter $\lambda$ are controlled simultaneously to maintain the Shannon entropy of the working substance constant. As has been derived in the supplementary material of Ref.~\cite{Martinez15}, the probability distribution function of the heat input through the Brownian isentropic process in the underdamped case (as in our experiment~\cite{Martinez16}) is given by
\begin{equation}
	P_{Q_{\text{ise}}^{\text{Brow}}}(q) = 
\begin{cases}
	\dfrac{1}{k_{\rm B}(T_{\rm{init}}+T_{\rm{fin}})} \mathrm{e}^{ -\beta_{\rm{fin}} \left(q - \avg{W_{\text{ise}}^{\text{Brow}}} \right) } &\text{~~~~for~~} q - \avg{W_{\text{ise}}^{\text{Brow}}} \ge 0, \\[1em]
	\dfrac{1}{k_{\rm B}(T_{\rm{init}}+T_{\rm{fin}})} \mathrm{e}^{ \beta_{\rm{init}} \left(q - \avg{W_{\text{ise}}^{\text{Brow}}} \right) } &\text{~~~~for~~} q - \avg{W_{\text{ise}}^{\text{Brow}}} < 0.
\end{cases}
\end{equation}
Here, $\avg{W_{\text{ise}}^{\text{Brow}}}$ is the mean value of the work output through the Brownian isentropic process.

Now, we calculate the moments of work output in the true adiabatic process and heat input in the Brownian isentropic process. First, we consider the adiabatic expansion stroke $1^+ \rightarrow 2^-$ (see Fig.~\ref{fig:carnot_suppl}) with $T_\text{init} = T_{\rm h}$,\, $T_\text{fin} = T_{\rm c}$, and $T_\text{init} > T_\text{fin}$. The distribution function of work output $W^{\text{true}}_{1 \rightarrow 2}$ through the true adiabatic process, $P_{W^{\text{true}}_{1 \rightarrow 2} }(w)$, and that of the heat input $Q_{1 \rightarrow 2}^{\text{Brow}}$ through the corresponding Brownian isentropic process, $P_{Q_{1 \rightarrow 2}^{\text{Brow}}} (q)$, are as follows:
\begin{align}
	P_{W^{\text{true}}_{1 \rightarrow 2} }(w) &=
	\begin{cases}
		\dfrac{1}{k_{\rm B} \Delta  T} \mathrm{e}^{-\frac{1}{k_{\rm B} \Delta T} w } &\text{~~for~~} w \ge 0, \\[1em]
		0   &\text{~~for~~} w < 0,
	\end{cases}\\\nonumber\\
	P_{Q_{1 \rightarrow 2}^{\text{Brow}}} (q) &= 
	\begin{cases}
		\dfrac{1}{k_{\rm B}(T_{\rm h}+T_{\rm c})} \mathrm{e}^{ -\frac{1}{k_{\rm B} T_{\rm c}}\left(q - \avg{W_{1 \rightarrow 2}^{\text{Brow}}} \right) }  &\text{~~~~for~~} q - \avg{W_{1 \rightarrow 2}^{\text{Brow}}} \ge 0, \\[1em]
		\dfrac{1}{k_{\rm B}(T_{\rm h}+T_{\rm c})} \mathrm{e}^{ -\frac{1}{k_{\rm B} T_{\rm h}}\left(-q + \avg{W_{1 \rightarrow 2}^{\text{Brow}}} \right) }  &\text{~~~~for~~} q - \avg{W_{1 \rightarrow 2}^{\text{Brow}}} < 0,
	\end{cases}
\end{align}
with $\Delta T \equiv T_{\rm h}-T_{\rm c}$. From these distribution functions, the mean values of $W^{\text{true}}_{1 \rightarrow 2}$ and $Q_{1 \rightarrow 2}^{\text{Brow}}$ are calculated as
\begin{equation}
	\avg{W^{\text{true}}_{1 \rightarrow 2}} = \int_{0}^{\infty} dw \, w \, P_{W^{\text{true}}_{1 \rightarrow 2} }(w) = k_{\rm B} \Delta T   \label{eq:worktrue_1}
\end{equation}
and
\begin{equation}
	\avg{Q_{1 \rightarrow 2}^{\text{Brow}}} = \int_{-\infty}^{\infty} dq \, q \, P_{Q_{1 \rightarrow 2}^{\text{Brow}}} (q) = -k_{\rm B} \Delta T +  \avg{W_{1 \rightarrow 2}^{\text{Brow}}} . \label{eq:heatB_1}
\end{equation}
Thus, the relation between the mean values of $W^{\text{true}}_{1 \rightarrow 2} $ and $Q_{1 \rightarrow 2}^{\text{Brow}}$ reads
\begin{equation}
	\avg{W^{\text{true}}_{1 \rightarrow 2}}  = 	- \avg{Q_{1 \rightarrow 2}^{\text{Brow}}} +  \avg{W_{1 \rightarrow 2}^{\text{Brow}}} .
\end{equation}
Similarly, for the second moments of $W^{\text{true}}_{1 \rightarrow 2} $ and $Q_{1 \rightarrow 2}^{\text{Brow}}$, we obtain 
\begin{equation}
	\avg{\left(W^{\text{true}}_{1 \rightarrow 2}\right)^2} = 2(k_{\rm B} \Delta T)^2 \label{eq:worktrue_2}
\end{equation}
and
\begin{equation}
	\avg{\left( Q_{1 \rightarrow 2}^{\text{Brow}} \right)^2} =  2 k_{\rm B}^2 T_{\rm h} T_{\rm c} + 2 (k_{\rm B} \Delta T)^2 - 2 k_{\rm B} \Delta T \avg{W_{1 \rightarrow 2}^{\text{Brow}}} + \avg{W_{1 \rightarrow 2}^{\text{Brow}}} ^2 .  \label{eq:heatB_2}
\end{equation}
 From Eqs.~(\ref{eq:worktrue_1}), (\ref{eq:heatB_1}),(\ref{eq:worktrue_2}), and (\ref{eq:heatB_2}), we obtain the relation between the variances of $W^{\text{true}}_{1 \rightarrow 2} $ and $Q_{1 \rightarrow 2}^{\text{Brow}}$:
 \begin{equation}
 	\avg{\left( \Delta W^{\text{true}}_{1 \rightarrow 2}  \right)^2} = \avg{\left( \Delta Q_{1 \rightarrow 2}^{\text{Brow}} \right)^2} - 2 k_{\rm B}^2 T_{\rm h} T_{\rm c}. \label{eq:rel2ndmom_1}
 \end{equation}
In the same way, we obtain the relations for the 3rd and 4th moments:
\begin{equation}
	\avg{\left( \Delta W^{\text{true}}_{1 \rightarrow 2}  \right)^3} = \avg{\left( \Delta Q_{1 \rightarrow 2}^{\text{Brow}} \right)^3} + 6 \, k_{\rm B}^3 T_{\rm h} T_{\rm c} \Delta T 
\end{equation}
and
\begin{equation}
	\avg{\left( \Delta W^{\text{true}}_{1 \rightarrow 2}  \right)^4} = \avg{\left( \Delta Q_{1 \rightarrow 2}^{\text{Brow}} \right)^4} - 12 \, k_{\rm B}^4 T_{\rm h} T_{\rm c} (3 \Delta T^2 + 2 T_{\rm h} T_{\rm c} ).
\end{equation}

For the adiabatic compression stroke $3^+ \rightarrow 4^-$ (see Fig.~\ref{fig:carnot_suppl}), following the same procedure, we obtain the similar relations between $W^{\text{true}}_{3 \rightarrow 4} $ and $Q_{3 \rightarrow 4}^{\text{Brow}}$:
\begin{align}
	\avg{W^{\text{true}}_{3 \rightarrow 4}}  =&\, 	- \avg{Q_{3 \rightarrow 4}^{\text{Brow}}} +  \avg{W_{3 \rightarrow 4}^{\text{Brow}}} ,\\[1em]
    \avg{\left( \Delta W^{\text{true}}_{3 \rightarrow 4}  \right)^2} =&\, \avg{\left( \Delta Q_{3 \rightarrow 4}^{\text{Brow}} \right)^2} - 2 k_{\rm B}^2 T_{\rm h} T_{\rm c}, \label{eq:rel2ndmom_2}\\[1em]
    \avg{\left( \Delta W^{\text{true}}_{3 \rightarrow 4}  \right)^3} =&\, \avg{\left( \Delta Q_{3 \rightarrow 4}^{\text{Brow}} \right)^3} + 6 \, k_{\rm B}^3 T_{\rm h} T_{\rm c} \Delta T ,\\[1em]
    \avg{\left( \Delta W^{\text{true}}_{3 \rightarrow 4}  \right)^4} =&\, \avg{\left( \Delta Q_{3 \rightarrow 4}^{\text{Brow}} \right)^4} - 12 \, k_{\rm B}^4 T_{\rm h} T_{\rm c} (3 \Delta T^2 + 2 T_{\rm h} T_{\rm c} ).
\end{align}

\medskip
\noindent \textit{Evaluation of $\eta^{(2)}$ from experimental data.}

Now, we discuss the evaluation of $\eta^{(2)}$ using experimental data of the Brownian Carnot engine~\cite{Martinez16}. 
For the Carnot cycle in the quasistatic limit, the variance of total work output $W$ through one cycle comes solely from the true adiabatic strokes:
\begin{equation}
    \avg{\Delta W^2} = \avg{(\Delta W^{\text{true}}_{1 \rightarrow 2})^2} + \avg{(\Delta W^{\text{true}}_{3 \rightarrow 4})^2} = \sum_{i} \avg{(\Delta W^{\text{true}}_i)^2},\label{eq:dw2_suppl}
\end{equation}
where $i=1 \rightarrow 2$ ($i=3 \rightarrow 4$) for the adiabatic compression (expansion).
To evaluate $\avg{\Delta W^2}$ from the experimental data, we use the relations (\ref{eq:rel2ndmom_1}) and (\ref{eq:rel2ndmom_2}) between the work variance $\avg{(\Delta W^{\text{true}}_i)^2}$ of the true adiabatic strokes and the heat variance $\avg{\left( \Delta Q_i^{\text{Brow}} \right)^2}$ of the corresponding Brownian isentropic strokes. In the analysis, we identify the sample variance $\avg{\Delta Q_i^2}_{\text{exp}}$ of the experimental data of heat in the Brownian isentropic strokes with $\avg{\left( \Delta Q_i^{\text{Brow}} \right)^2}$. In addition, since the actual effective temperature of the working substance in the isothermal strokes generally deviates from the temperature of the bath in the non-quasistatic regime, we use the time average of the effective temperature of the working substance during the hot (cold) isothermal stroke in the experiment, denoted by $T_{\rm h}^{\text{eff}}$ ($T_{\rm c}^{\text{eff}}$), for $T_{\rm h}$ ($T_{\rm c}$) in these relations. (Note that $T_{\rm h}^{\text{eff}}$ and $T_{\rm c}^{\text{eff}}$ converge to the temperatures of the baths in the quasistatic limit.) As a consequence, we employ the following relation:
\begin{equation}
\avg{(\Delta W^{\text{true}}_i)^2} = \avg{\Delta Q_i^2}_{\text{exp}} - 2k_{\rm B}^2 T_{\rm h}^{\text{eff}} T_{\rm c}^{\text{eff}}.\label{eq:rel2ndmom_exp}
\end{equation}
Hereafter, for the sample variance of some quantity $O$ over the experimental realizations, we write $\avg{\Delta O^2}_{\text{exp}}$ for clarity.

The variance $\avg{\Delta Q_{i}^2}_{\text{exp}}$ of heat can be decomposed into the variance $\avg{\Delta Q_{{\rm k},i}^2}_{\text{exp}}$ of the kinetic energy and the variance $\avg{\Delta Q_{{\rm p},i}^2}_{\text{exp}}$ of the potential energy: $\avg{\Delta Q_{i}^2}_{\text{exp}} = \avg{\Delta Q_{{\rm k},i}^2}_{\text{exp}} + \avg{\Delta Q_{{\rm p},i}^2}_{\text{exp}}$. Here, $dQ_{{\rm p},i} = -\partial_x V(x,\lambda) dx$ is obtained from the trajectory $x(t)$ realized in the experiment. Since the relaxation time of the velocity of the Brownian particle is much shorter than the other time scales, we can assume that the kinetic energy is always distributed as in equilibrium with the environment. Thus, we obtain $\avg{\Delta Q_{{\rm k},i}^2}_{\text{exp}} = \avg{\Delta E_{\text{k},\text{init}}^2}_{\text{exp}} + \avg{\Delta E_{\text{k},\text{fin}}^2}_{\text{exp}} = \frac{1}{2}k_{\rm B}^2 (T_{\text{init}}^2 + T_{\text{fin}}^2)$ from the variance of the kinetic energy, $E_{\text{k},\text{init}}$ and $E_{\text{k},\text{fin}}$, at two ends of each Brownian isentropic stroke, where $T_{\text{init}} = T_{\rm h}$ and $T_{\text{fin}} = T_{\rm c}$ for $i= 1 \rightarrow 2$ ($T_{\text{init}} = T_{\rm c}$ and $T_{\text{fin}} = T_{\rm h}$ for $i= 3 \rightarrow 4$). Consequently, $\avg{\Delta Q_i^2}_{\text{exp}}$ for the Brownian isentropic stroke is given by 
\begin{equation}
    \avg{\Delta Q_i^2}_{\text{exp}} = \avg{\Delta Q_{{\rm p},i}^2}_{\text{exp}} + \frac{1}{2}k_{\rm B}^2 (T_{\rm c}^2 + T_{\rm h}^2).\label{eq:dq2Bise_exp}
\end{equation}
Similarly, the variance of heat during the hot isothermal stroke ($2 \rightarrow 3$), where $T_{\text{init}} = T_{\text{fin}} = T_{\rm h}$, is given by
\begin{equation}
    \avg{\Delta Q_{2 \rightarrow 3}^2}_{\text{exp}} = \avg{\Delta Q_{{\rm p},2 \rightarrow 3}^2}_{\text{exp}} + k_{\rm B}^2 T_{\rm h}^2.\label{eq:dq2Bisoth_exp}
\end{equation}
From Eqs.~(\ref{eq:dw2_suppl}), (\ref{eq:rel2ndmom_exp}), (\ref{eq:dq2Bise_exp}), and (\ref{eq:dq2Bisoth_exp}), we finally obtain $\eta^{(2)}$ as
\begin{equation}
    \eta^{(2)} = \frac{\avg{\Delta Q_{{\rm p},1 \rightarrow 2}^2}_{\text{exp}} + \avg{\Delta Q_{{\rm p},3 \rightarrow 4}^2}_{\text{exp}} - 4 k_{\rm B}^2 T_{\rm h}^{\text{eff}}T_{\rm c}^{\text{eff}} + k_{\rm B}^2 (T_{\rm c}^2 + T_{\rm h}^2)}
    {\avg{\Delta Q_{{\rm p},2 \rightarrow 3}^2}_{\text{exp}} + k_{\rm B}^2 T_{\rm h}^2},\label{eq:eta2_exp_suppl}
\end{equation}
which we evaluate by using the experimental data of $\avg{\Delta Q_{{\rm p},i}^2}_{\text{exp}}$, $T_{\rm h}^{\text{eff}}$, and $T_{\rm c}^{\text{eff}}$.

\medskip
\noindent \textit{Evaluation of sampling error}

Due to the finite number of samples (i.e., finite number of realizations of trajectories), the sample variance $\avg{\Delta Q_{\rm p}^2}_{\text{exp}}$ of $Q_{\rm p}$ obtained from the sample trajectories can deviate from the theoretical value of the variance $\avg{\left( \Delta Q_{\rm p}^{\text{Brow}} \right)^2}$. To calculate the error bars shown in Fig.~\ref{fig:experiment}, we simulate this deviation due to the finite number of trajectories realized in the experiment as follows.

In the quasistatic limit, we assume $Q_{\rm p} = E_{{\rm p},\text{fin}} - E_{{\rm p},\text{init}}$, where the probability density function of the potential energy $E_{\text{p},j}$ at node $j$ is given by $P(E_{\text{p}, j}) = (\pi k_{\rm{B}} T_j E_{\text{p}, j})^{-1/2} \exp{(-E_{\text{p}, j}/k_{\rm B}T_j)} $. We evaluate the error in $\avg{\Delta Q_{\rm p}^2}_{\text{exp}}$ caused by the finite number of samples in the experiment by a Monte-Carlo simulation based on this $P(E_{\text{p}, j})$. From the errors of $\avg{\Delta Q_{{\rm p},1 \rightarrow 2}^2}_{\text{exp}}$, $\avg{\Delta Q_{{\rm p},2 \rightarrow 3}^2}_{\text{exp}}$, and $\avg{\Delta Q_{{\rm p},3 \rightarrow 4}^2}_{\text{exp}}$ obtained by the simulation, the error of $\eta^{(2)}$ is finally obtained through the error propagation formula for Eq.~(\ref{eq:eta2_exp_suppl}).

\bigskip
\noindent \textbf{Data availability}\\
All the simulation data in this paper can be reproduced using the described methodology. The relevant experimental data are available upon reasonable request. Source data of Fig.~\ref{fig:experiment} are provided with this paper as Supplementary Data 1.

\bigskip
\noindent \textbf{Code availability}\\
The codes used to generate the figures are available upon reasonable request.

\bigskip
\noindent \textbf{Acknowledgement}\\
We thank Yuki Izumida, Peter Talkner, and B. Prasanna Venkatesh for helpful discussions and comments. This work is supported by NSF of China (Grant No.~12375039, No.~11975199, and No.~11674283), by the Zhejiang Provincial Natural Science Foundation Key Project (Grant No.~LZ19A050001), by the Fundamental Research Funds for the Central Universities (2017QNA3005, 2018QNA3004), and by the Zhejiang University 100 Plan. K.I. is also supported by MEXT Quantum Leap Flagship Program (MEXT Q-LEAP) (Grant No. JPMXS0118067394 and No. JPMXS0120319794). I.A.M. is supported by MSCA-IF NEQLIQ-101030465. \'E.R. acknowledges financial support from PNRR MUR project PE0000023-NQSTI. K.I. and G.-H.X. contributed equally.

\bigskip
\noindent \textbf{Author contributions}\\
G.W. designed the project. K.I., G.-H.X., C.J. and G.W. conducted the theoretical study, and I.A.M., \'E.R. and R.A.R.-A. conducted the experiment and its data analysis. K.I. and G.-H.X. contributed equally.

\bigskip
\noindent \textbf{Competing interests}\\
The authors declare no competing interests.

\end{document}


\title{--- Supplementary Information --- \\
  Universal relations and bounds for fluctuations in quasistatic small heat engines}

\author{Kosuke Ito}
\affiliation{School of Physics and Zhejiang Institute of Modern Physics, Zhejiang University, Hangzhou, Zhejiang 310027, China}
\affiliation{Graduate School of Engineering Science, Osaka University, 1-3 Machikaneyama, Toyonaka, Osaka 560-8531, Japan}
\affiliation{Center for Quantum Information and Quantum Biology, Osaka University, 1-2 Machikaneyama, Toyonaka, Osaka 560-8531, Japan}
\author{Guo-Hua Xu}
\affiliation{School of Physics and Zhejiang Institute of Modern Physics, Zhejiang University, Hangzhou, Zhejiang 310027, China}
\author{Chao Jiang}
\affiliation{School of Physics and Zhejiang Institute of Modern Physics, Zhejiang University, Hangzhou, Zhejiang 310027, China}

\author{\'Edgar Rold\'an} 
\affiliation{ICTP -- The Abdus Salam International Centre for Theoretical Physics, Strada Costiera 11, 34151 Trieste, Italy}

\author{Ra\'ul~A.~Rica-Alarc\'on}   
\affiliation{Universidad de Granada, Department of Applied Physics, 18071 Granada, Spain}
\affiliation{Universidad de Granada, Nanoparticles Trapping Laboratory and Research Unit ``Modeling Nature" (MNat), 18071 Granada, Spain}

\author{Ignacio~A. Mart\'inez}  
\affiliation{Department of Electronics and Information Systems, Ghent University, Technologiepark-Zwijnaarde 126, 9052 Ghent, Belgium}

\author{Gentaro Watanabe}\email{gentaro@zju.edu.cn}
\affiliation{School of Physics and Zhejiang Institute of Modern Physics, Zhejiang University, Hangzhou, Zhejiang 310027, China}
\affiliation{Zhejiang Province Key Laboratory of Quantum Technology and Device, Zhejiang University, Hangzhou, Zhejiang 310027, China}


\maketitle
\tableofcontents

\subsection{Supplementary Note 1: An example giving $\eta^{(2)} > \eta_{\rm C}^{(2)}$}

\begin{figure}[t]
\centering
\includegraphics[width=0.33 \columnwidth]{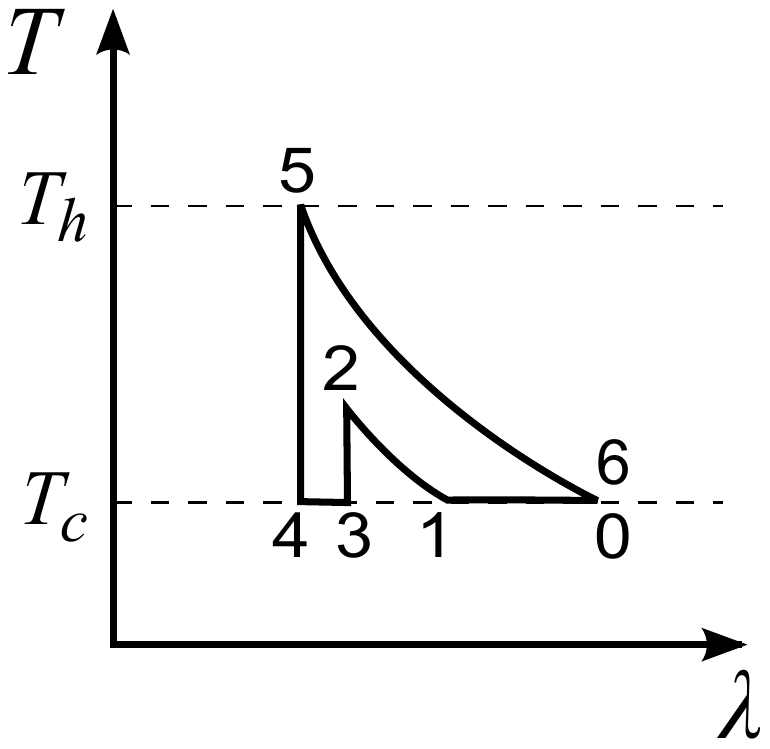}
\caption{\textbf{An example of a cycle on the $T$-$\lambda$ plane whose $\eta^{(2)}$ can be larger than that of the Carnot cycle $\eta_{\rm C}^{(2)}$.} In this cycle, a quasistatic adiabatic compression stroke $1^+ \rightarrow 2^-$ is followed by a cold isochoric stroke $2^+ \rightarrow 3$ instead of a hot one.
}
\label{fig:exception}
\end{figure}

The relation $\eta^{(2)} \le \eta_{\rm C}^{(2)}$ holds provided all the quasistatic adiabatic expansion strokes are preceded by a stroke with a hot bath and all the quasistatic adiabatic compression strokes are followed by a stroke with a hot bath. However, if it is not the case, this relation is not guaranteed. Here we provide such an example.

Figure \ref{fig:exception} shows the $T$-$\lambda$ diagram of the cycle which we shall consider. At point $0$, we set the external parameter at $\lambda_0$ and take a randomly chosen microstate from the canonical ensemble for $H_{\lambda_0}$ at temperature $T_{\rm c}$. Starting with this initial state, we perform the following six strokes.\, $(0^+ \rightarrow 1^-)$: Quasistatic isothermal compression at $T_{\rm c}$.\, $(1^+ \rightarrow 2^-)$: Quasistatic adiabatic compression.\, $(2^+ \rightarrow 3)$: Cold isochore.\, $(3 \rightarrow 4^-)$: Quasistatic isothermal compression at $T_{\rm c}$.\, $(4^+ \rightarrow 5^-)$: Hot isochore.\, $(5^+ \rightarrow 6^-)$: Quasistatic adiabatic expansion. The final point $6^-$ is equivalent to the initial point $0$, and the cycle is closed. (Regarding point $3$, there is no need to distinguish between $3^-$ and $3^+$ since the engine is kept in contact with the cold heat bath throughout the process $2^+ \rightarrow 3 \rightarrow 4^-$.) In this example, the quasistatic adiabatic compression stroke $2^+ \rightarrow 3^-$ is not followed by a stroke with a hot bath.

We take a working substance whose number of states is in the form of
\begin{align}
  I_\lambda(E) \equiv \int d\Gamma\,\, \theta\left(E - H_{\lambda}(\Gamma)\right) = f(\lambda)\, E^\alpha\,,
\end{align}
with an arbitrary function $f$ and a real constant $\alpha$. In this case, the density of state $g_\lambda(E)$ reads
\begin{align}
  g_\lambda(E) \equiv \frac{\partial I_\lambda(E)}{\partial E} = \int d\Gamma\,\, \delta\left(E - H_{\lambda}(\Gamma)\right) = \alpha\, f(\lambda)\, E^{\alpha - 1}\,.\label{eq:g_suppl}
\end{align}
Then, the canonical distribution $P_{\beta,\, \lambda}^{\rm eq}$ at the inverse temperature $\beta \equiv 1/k_{\rm B}T$ and the external parameter $\lambda$ is
\begin{align}
  P_{\beta,\, \lambda}^{\rm eq}(E) = \frac{g_\lambda(E)}{Z_{\beta,\, \lambda}} \mathrm{e}^{-\beta E} = \frac{\beta^\alpha\, E^{\alpha - 1}}{\Gamma(\alpha)} \mathrm{e}^{-\beta E}\,,\label{eq:canonical_suppl}
\end{align}
with $Z_{\beta,\, \lambda}$ being the partition function given by
\begin{align}
  Z_{\beta,\, \lambda} = \int_0^\infty dE\,\, g_\lambda(E)\, \mathrm{e}^{-\beta E} = \alpha\, \Gamma(\alpha)\, f(\lambda)\, \beta^{-\alpha}\,,\label{eq:partition_suppl}
\end{align}
where $\Gamma(\alpha) \equiv \int_0^\infty dx\, x^{\alpha - 1} \mathrm{e}^{-x}$ is the gamma function.

For the canonical distribution given by Eq.~(\ref{eq:canonical_suppl}), the average of the internal energy $E_i$ and of its square $E_i^2$ at point $i$ with temperature $T_i$ and parameter $\lambda_i$ can be calculated as
\begin{align}
  \avg{E_i} &= \int_0^\infty dE\,\, E\, P_{\beta_i,\, \lambda_i}^{\rm eq}(E) = k_{\rm B}T_i \alpha\,,\label{eq:ei_suppl}\\
  \avg{E_i^2} &= \int_0^\infty dE\,\, E^2\, P_{\beta_i,\, \lambda_i}^{\rm eq}(E) = (k_{\rm B}T_i)^2\, (\alpha+1)\, \alpha\,.
\end{align}
Therefore, the variance $\avg{\Delta E_i^2}$ of $E_i$ reads
\begin{align}
  \avg{\Delta E_i^2} = \avg{E_i^2} - \avg{E_i}^2 = (k_{\rm B}T_i)^2 \alpha\,.\label{eq:deltaesq_suppl}
\end{align}

Heat input $Q_{\rm h}$ from the hot heat bath is done only in the stroke $4^+ \rightarrow 5^-$ throughout the cycle: $Q_{\rm h} = Q_{4 \rightarrow 5}$ and $\Delta Q_{\rm h} \equiv Q_{\rm h} - \avg{Q_{\rm h}} = \Delta Q_{4 \rightarrow 5}$. Therefore, according to Table I in the main paper and Eq.~(\ref{eq:deltaesq_suppl}), the variance of $Q_{\rm h}$ can be written as
\begin{align}
  \avg{\Delta Q_{\rm h}^2} = \avg{\Delta Q_{4 \rightarrow 5}^2} = \avg{\Delta E_4^2} + \avg{\Delta E_5^2} = k_{\rm B}^2 (T_{\rm c}^2 + T_{\rm h}^2)\, \alpha\,.\label{eq:deltaqhsq_suppl}
\end{align}
Next, work output $W$ through the whole cycle is given by
\begin{align}
  W = W_{0 \rightarrow 1} + W_{1 \rightarrow 2} + W_{2 \rightarrow 3} + W_{3 \rightarrow 4} + W_{4 \rightarrow 5} + W_{5 \rightarrow 6} = W_{0 \rightarrow 1} + W_{1 \rightarrow 2} + W_{3 \rightarrow 4} + W_{5 \rightarrow 6}\,
\end{align}
because no work is done in the isochoric strokes: $W_{2 \rightarrow 3} = W_{4 \rightarrow 5} = 0$. In addition, since the fluctuations of work outputs $W_{0 \rightarrow 1}$ and $W_{3 \rightarrow 4}$ through the quasistatic isothermal strokes are negligible, we get
\begin{align}
  \Delta W \equiv W - \avg{W} = \Delta W_{1 \rightarrow 2} + \Delta W_{5 \rightarrow 6}\,.
\end{align}
Using Eq.~(\ref{eq:deltaesq_suppl}) and the expression of the work fluctuation through a quasistatic adiabatic stroke given in Table I of the main paper, we get $\avg{\Delta W_{1 \rightarrow 2}^2} = k_{\rm B}^2 (T_{\rm c} - T_2)^2 \alpha$ and $\avg{\Delta W_{5 \rightarrow 6}^2} = k_{\rm B}^2 (T_{\rm h} - T_{\rm c}) \alpha$. Since $W_{1 \rightarrow 2}$ and $W_{5 \rightarrow 6}$ are independent random variables, the variance $\avg{\Delta W^2}$ of $W$ reads
\begin{align}
  \avg{\Delta W^2} = \avg{\Delta W_{1 \rightarrow 2}^2} + \avg{\Delta W_{5 \rightarrow 6}^2} = k_{\rm B}^2 (T_{\rm c} - T_2)^2 \alpha + k_{\rm B}^2 (T_{\rm h} - T_{\rm c})^2 \alpha\,.\label{eq:deltawsq_suppl}
\end{align}
From Eqs.~(\ref{eq:deltaqhsq_suppl}) and (\ref{eq:deltawsq_suppl}), we finally obtain
\begin{align}
  \eta^{(2)} \equiv \frac{\avg{\Delta W^2}}{\avg{\Delta Q_{\rm h}^2}} = \frac{(T_{\rm h} - T_{\rm c})^2 + (T_2 - T_{\rm c})^2}{T_{\rm c}^2 + T_{\rm h}^2}\,.
\end{align}
This $\eta^{(2)}$ is apparently an increasing function of $T_2$, and it becomes $\eta^{(2)} > \eta_{\rm C}^{(2)}$ when
\begin{align}
  T_2 > T_{\rm c} \left( 2 - \frac{T_{\rm c}}{T_{\rm h}} \right) \equiv T_*\,.
\end{align}
Note that $T_*$ is always smaller than $T_{\rm h}$ provided $T_{\rm c}/T_{\rm h} < 1$.

\bigskip
\subsection{Supplementary Note 2: Another example giving $\eta_{\rm C}^{(2)}$ and $\xi_{\rm C}^{(2)}$}

In the main paper, we have shown that the Carnot cycle gives upper bounds $\eta_{\rm C}^{(2)}$ and $\xi_{\rm C}^{(2)}$ for $\eta^{(2)} \equiv \avg{\Delta W^2}/\avg{\Delta Q_{\rm h}^2}$ and $\xi^{(2)} \equiv \avg{\Delta W^2}/\Delta Q^{(2)}$, respectively. However, the Carnot cycle is not the unique case which gives these bounds. Here we provide an example whose $\eta^{(2)}$ and $\xi^{(2)}$ are the same as those for the Carnot cycle.

\begin{figure}[t]
\centering
\includegraphics[width=0.55 \columnwidth]{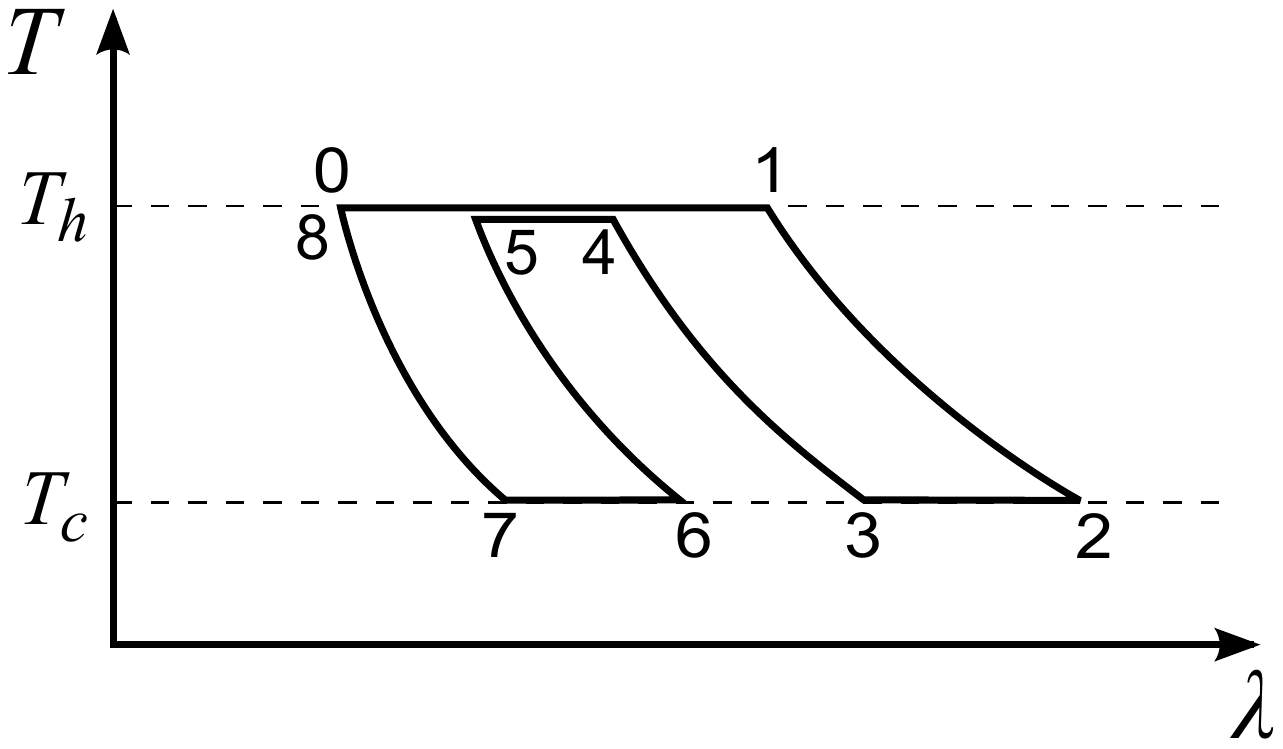}
\caption{\textbf{An example of a cycle whose $\eta^{(2)}$ and $\xi^{(2)}$ is the same as those of the Carnot cycle.} The temperature versus external control parameter ($T$-$\lambda$) diagram of this cycle working with a hot heat bath at temperature $T_{\rm h}$ and a cold one at $T_{\rm c}$. 
}
\label{fig:doublecarnot}
\end{figure}

The example on the temperature versus external parameter ($T$-$\lambda$) plane is shown in Fig.~\ref{fig:doublecarnot}. As can be seen from this figure, this cycle can be regarded as a combination of two Carnot cycles. At point $0$, we set the external parameter at $\lambda_0$ and take a randomly chosen microstate from the canonical ensemble for $H_{\lambda_0}$ at temperature $T_{\rm h}$. Starting with this initial state, we perform the following eight strokes in numerical order of the points: two quasistatic isothermal strokes at $T_{\rm h}$ [$(0^+ \rightarrow 1^-)$ and $(4^+ \rightarrow 5^-)$], those at $T_{\rm c}$ [$(2^+ \rightarrow 3^-)$ and $(6^+ \rightarrow 7^-)$], two quasistatic adiabatic expansion strokes [$(1^+ \rightarrow 2^-)$ and $(5^+ \rightarrow 6^-)$], and two quasistatic adiabatic compression strokes [$(3^+ \rightarrow 4^-)$ and $(7^+ \rightarrow 8^-)$]. The final point $8^-$ is equivalent to the initial point $0$, and the cycle is closed. Here, the superscripts ``$-$'' and ``$+$'' respectively represent just before and after making/removing a thermal contact with a heat bath when finishing/starting the quasistatic adiabatic stroke.

Heat input $Q_{\rm h}$ from the hot heat bath throughout this cycle is given by
\begin{align}
  Q_{\rm h} = Q_{0 \rightarrow 1} + Q_{4 \rightarrow 5}\,,
\end{align}
where $Q_{0 \rightarrow 1}$ and $Q_{4 \rightarrow 5}$ are heat input through the quasistatic isothermal strokes $(0^+ \rightarrow 1^-)$ and $(4^+ \rightarrow 5^-)$ at temperature $T_{\rm h}$, respectively:
\begin{align}
  Q_{0 \rightarrow 1} &= E_1^- - E_0^+ + W_{0 \rightarrow 1}\,,\\
  Q_{4 \rightarrow 5} &= E_5^- - E_4^+ + W_{4 \rightarrow 5}\,.
\end{align}
Here, $E_i^\pm$ represents the internal energy of the working substance at point $i^\pm$, and $W_{0 \rightarrow 1}$ and $W_{4 \rightarrow 5}$ are work output through the strokes $(0^+ \rightarrow 1^-)$ and $(4^+ \rightarrow 5^-)$, respectively. Since the fluctuations of $W_{0 \rightarrow 1}$ and $W_{4 \rightarrow 5}$ are negligible, we get
\begin{align}
  \Delta Q_{\rm h} &\equiv Q_{\rm h} - \avg{Q_{\rm h}}\,\nonumber\\
  &= (\Delta E_1^- - \Delta E_0^+) + (\Delta E_5^- - \Delta E_4^+)\,.
\end{align}
with $\Delta E_i^\pm \equiv E_i^\pm - \avg{E_i^\pm}$.

Similarly, heat output $Q_{\rm c}$ to the cold heat bath throughout the cycle is
\begin{align}
  Q_{\rm c} = -(Q_{2 \rightarrow 3} + Q_{6 \rightarrow 7})\,,
\end{align}
where $Q_{2 \rightarrow 3}$ and $Q_{6 \rightarrow 7}$ are heat input through the quasistatic isothermal strokes $(2^+ \rightarrow 3^-)$ and $(6^+ \rightarrow 7^-)$ at $T_{\rm c}$, respectively:
\begin{align}
  Q_{2 \rightarrow 3} = E_3^- - E_2^+ + W_{2 \rightarrow 3}\,,\\
  Q_{6 \rightarrow 7} = E_7^- - E_6^+ + W_{6 \rightarrow 7}\,.
\end{align}
Since the fluctuations of $W_{2 \rightarrow 3}$ and $W_{6 \rightarrow 7}$ are negligible, we get
\begin{align}
  \Delta Q_{\rm c} \equiv Q_{\rm c} - \avg{Q_{\rm c}} = - \left[ (\Delta E_3^- - \Delta E_2^+) + (\Delta E_7^- - \Delta E_6^+) \right]\,.
\end{align}
Thus the variance $\avg{\Delta Q_{\rm c}^2}$ of $Q_{\rm c}$ can be written as
\begin{align}
  \avg{\Delta Q_{\rm c}^2} &= \avg{\left[ (\Delta E_3^- - \Delta E_2^+) + (\Delta E_7^- - \Delta E_6^+) \right]^2}\,\nonumber\\
  &= \avg{\left[ (\Delta E_3^+ - \Delta E_2^-) + (\Delta E_7^+ - \Delta E_6^-) \right]^2}\nonumber\\
  &= \left(\frac{T_{\rm c}}{T_{\rm h}}\right)^2 \avg{\left[ (\Delta E_4^- - \Delta E_1^+) + (\Delta E_8^- - \Delta E_5^+) \right]^2}\,\nonumber\\
  &= \left(\frac{T_{\rm c}}{T_{\rm h}}\right)^2 \avg{\left[ (\Delta E_4^+ - \Delta E_1^-) + (\Delta E_0^+ - \Delta E_5^-) \right]^2}\,\nonumber\\
  &= \left(\frac{T_{\rm c}}{T_{\rm h}}\right)^2 \avg{\Delta Q_{\rm h}^2}\,.\label{eq:deltaqcsq_suppl}
\end{align}
From the first to the second line, we have used the relation $E_i^+ = E_i^-$, which allows us to replace $\Delta E_i^+$ by $\Delta E_i^-$. The relation $E_i^+ = E_i^-$ is a direct consequence of the first law of thermodynamics together with the fact that the work for making and removing a thermal contact with a heat bath is neglibible \cite{Sekimotobook10suppl,Sekimotobookjpsuppl}. From the second to the third line, we have used the relation given by Eq.~(1) of the main paper for the quasistatic adiabatic strokes $(1^+ \rightarrow 2^-)$, $(3^+ \rightarrow 4^-)$, $(5^+ \rightarrow 6^-)$, and $(7^+ \rightarrow 8^-)$ reading $E_2^- = (T_{\rm c}/T_{\rm h})\, E_1^+$, $E_3^+ = (T_{\rm c}/T_{\rm h})\, E_4^-$, $E_6^- = (T_{\rm c}/T_{\rm h})\, E_5^+$, and $E_7^+ = (T_{\rm c}/T_{\rm h})\, E_8^-$, respectively. Then, from the third to the fourth line, we have used the facts that $E_0$, $E_1$, $E_4$, $E_5$, and $E_8$ are independent, points $0$ and $8^-$ are equivalent, and $E_i^+ = E_i^-$.

From Eq.~(\ref{eq:deltaqcsq_suppl}), the sum $\Delta Q^{(2)}$ of the heat fluctuations over the separated sequances of heat exchanging strokes is given by
\begin{align}
  \Delta Q^{(2)} = \avg{\Delta Q_{\rm h}^2} + \avg{\Delta Q_{\rm c}^2} = \left[ 1 + \left(\frac{T_{\rm c}}{T_{\rm h}}\right)^2 \right] \avg{\Delta Q_{\rm h}^2}\,.\label{eq:deltaqsq_suppl}
\end{align}
(Since $Q_{\rm h}$ and $Q_{\rm c}$ are correlated, $\Delta Q^{(2)}$ is different from the variance $\avg{\Delta Q^2}$ of the total heat input $Q \equiv Q_{\rm h} - Q_{\rm c}$.)

Next, we consider the total work output $W$ through the whole cycle:
\begin{align}
  W = W_{0 \rightarrow 1} + W_{1 \rightarrow 2} + W_{2 \rightarrow 3} + W_{3 \rightarrow 4} + W_{4 \rightarrow 5} + W_{5 \rightarrow 6} + W_{6 \rightarrow 7} + W_{7 \rightarrow 8}\,.
\end{align}
Since the fluctuations of the work outputs $W_{0 \rightarrow 1}$, $W_{2 \rightarrow 3}$, $W_{4 \rightarrow 5}$, and $W_{6 \rightarrow 7}$ through the quasistatic isothermal strokes are negligible, we obtain
\begin{align}
  \Delta W \equiv W - \avg{W} = \Delta W_{1 \rightarrow 2} + \Delta W_{3 \rightarrow 4} + \Delta W_{5 \rightarrow 6} + \Delta W_{7 \rightarrow 8}\,.\label{eq:dwdouble_suppl}
\end{align}
Here, work through an adiabatic stroke is given by the difference between the internal energies of the initial and the final states of the stroke, which satisfy the adiabatic-reversible (AR) condition discussed in the main paper. For example, regarding the stroke $1^+ \rightarrow 2^-$, the work output $W_{1 \rightarrow 2}$ through this stroke can be written as
\begin{align}
  W_{1 \rightarrow 2} = E_1^+ - E_2^- = \left( 1 - \frac{T_{\rm c}}{T_{\rm h}} \right)\, E_1^+\,,
\end{align}
and thus
\begin{align}
  \Delta W_{1 \rightarrow 2} = \left( 1 - \frac{T_{\rm c}}{T_{\rm h}} \right)\, \Delta E_1^+\,.
\end{align}
Similarly, for the other quasistatic adiabatic strokes, we get $\Delta W_{3 \rightarrow 4} = [(T_{\rm c}/T_{\rm h}) - 1]\, \Delta E_4^-$, $\Delta W_{5 \rightarrow 6} = [1 - (T_{\rm c}/T_{\rm h})]\, \Delta E_5^+$, and $\Delta W_{7 \rightarrow 8} = [(T_{\rm c}/T_{\rm h}) - 1]\, \Delta E_8^-$. Substituting them into Eq.~(\ref{eq:dwdouble_suppl}), we obtain
\begin{align}
  \Delta W = \left(1 - \frac{T_{\rm c}}{T_{\rm h}}\right) \left[(\Delta E_1^+ - \Delta E_8^-) + (\Delta E_5^+ - \Delta E_4^-)\right]\,.
\end{align}
Since $E_0$, $E_1$, $E_4$, $E_5$, and $E_8$ are independent, points $0$ and $8^-$ are equivalent, and $E_i^+ = E_i^-$, we get
\begin{align}
  \avg{\Delta W^2} &= \left(1 - \frac{T_{\rm c}}{T_{\rm h}}\right)^2 \avg{\left[(\Delta E_1 - \Delta E_8^-) + (\Delta E_5 - \Delta E_4^-)\right]^2}\,\nonumber\\
  &= \left(1 - \frac{T_{\rm c}}{T_{\rm h}}\right)^2 \avg{\Delta Q_{\rm h}^2}\,.\label{eq:deltawsq1_suppl}
\end{align}

From Eqs.~(\ref{eq:deltaqsq_suppl}) and (\ref{eq:deltawsq1_suppl}), we finally obtain
\begin{align}
  \eta^{(2)} &\equiv \frac{\avg{\Delta W^2}}{\avg{\Delta Q_{\rm h}^2}} = \left( 1 - \frac{T_{\rm c}}{T_{\rm h}} \right)^2\,,\\
  \xi^{(2)} &\equiv \frac{\avg{\Delta W^2}}{\Delta Q^{(2)}} = \frac{\left[ 1 - (T_{\rm c}/T_{\rm h}) \right]^2}{1 + (T_{\rm c}/T_{\rm h})^2}\,,
\end{align}
which are the same as those for the Carnot cycle $\eta_{\rm C}^{(2)}$ and $\xi_{\rm C}^{(2)}$, respectively.